\documentclass{aa}
\usepackage{amssymb}
\usepackage{times}
\usepackage{mathptm}
\usepackage{graphics}
\usepackage{subfigure}
\usepackage{here}
\usepackage[dvips]{epsfig}
\usepackage{natbib}
\bibpunct{(}{)}{;}{a}{}{,}

\newcommand{\decdegmm}[2]{#1\mbox{$^\circ\mskip-6.6mu.\,$}#2}

\begin{document}     
\thesaurus
{04(04.19.1;03.13.2;08.23.1)}

\title{The stellar content of the Hamburg/ESO survey\thanks{Based on
    observations collected at the European Southern Observatory, La Silla and
    Paranal, Chile.}}
\subtitle{I. Automated selection of DA white dwarfs}

\author {N. Christlieb\inst{1}
\and L. Wisotzki\inst{2}
\and D. Reimers\inst{1}
\and D. Homeier\inst{3}  
\and D. Koester\inst{3}
\and U. Heber\inst{4}
}

\institute{Hamburger Sternwarte, Universit\"at Hamburg, Gojenbergsweg 112,
   D-21029 Hamburg, Germany; nchristlieb/dreimers@hs.uni-hamburg.de
\and Institut f\"ur Physik, Universit\"at Potsdam, Am Neuen Palais 10, 
  D-14469 Potsdam, Germany; lutz@astro.physik.uni-potsdam.de
\and Institut f\"ur Theoretische Physik und Astrophysik der
  Christian-Albrechts-Universit\"at Kiel, Leibnizstrasse 15, D-24098 Kiel,
   Germany; homeier/koester@astrophysik.uni-kiel.de
\and Dr. Remeis Sternwarte, Universit\"at Erlangen-N\"urnberg,
  Sternwartstrasse 7, D-96049 Bamberg; heber@sternwarte.uni-erlangen.de
}

\offprints{nchristlieb@hs.uni-hamburg.de}
\date{Received 12 September 2000 / Accepted 12 October 2000}
\titlerunning{Automated selection of DA white dwarfs}
\authorrunning{Christlieb et al.}
\maketitle

\begin{abstract}
  
  We describe automatic procedures for the selection of DA white dwarfs in the
  Hamburg/ESO objective-prism survey (HES). For this purpose, and the
  selection of other stellar objects (e.g., metal-poor stars and carbon
  stars), a flexible, robust algorithm for detection of stellar absorption and
  emission lines in the digital spectra of the HES was developed. Broad band
  ($U-B$, $B-V$) and narrow band (Str\"omgren $c_1$) colours can be derived
  directly from HES spectra, with precisions of $\sigma_{U-B}=0.092$\,mag;
  $\sigma_{B-V}=0.095$\,mag; $\sigma_{c_1}=0.15$\,mag.
  
  We describe simulation techniques that allow to convert model or
  slit spectra to HES spectra. These simulated objective-prism spectra are
  used to determine quantitative selection criteria, and for the study of
  selection functions. We present an atlas of simulated HES spectra of DA and
  DB white dwarfs.
  
  Our current selection algorithm is tuned to yield maximum efficiency of the
  candidate sample (minimum contamination with non-DAs). DA candidates are
  selected in the $B-V$ versus $U-B$ and $c_1$ versus $W_\lambda(\mbox{H}\beta
  +\mbox{H}\gamma+\mbox{H}\delta)$ parameter spaces. The contamination of the
  resulting sample with hot subdwarfs is expected to be as low as $\sim
  8$\,\%, while there is essentially no contamination with main sequence or
  horizontal branch stars. We estimate that with the present set of criteria,
  $\sim 80$\,\% of DAs present in the HES database are recovered. A yet higher
  degree of internal completeness could be reached at the expense of higher
  contamination. However, the external completeness is limited by additional
  losses caused by proper motion effects and the epoch differences between
  direct and spectral plates used in the HES.
\end{abstract}


\section{Introduction}

The Hamburg/ESO survey (HES) is an objective prism survey primarily targeting
bright quasars \citep{hespaperI,heshighlights,hespaperIII}. However, because
its spectral resolution is typically 15\,{\AA} FWHM at H$\gamma$, it is also
possible to efficiently select a variety of interesting \emph{stellar}
objects in the HES. These include, e.g., metal-poor halo stars, carbon stars,
cataclysmic variable stars, white dwarfs (WDs), subdwarf B stars (sdBs),
subdwarf O stars (sdOs), and field horizontal branch A- and B-type stars
\citep{Christlieb:2000}.  In a series of papers, we will report on the
development of quantitative selection procedures for the systematic
exploitation of the stellar content of the HES, and their application to the
digital HES data base. In this paper, we report on an automatic algorithm for
the selection of DA white dwarfs (DAs). 

\begin{figure*}[htbp]
  \begin{center}
    \epsfig{file=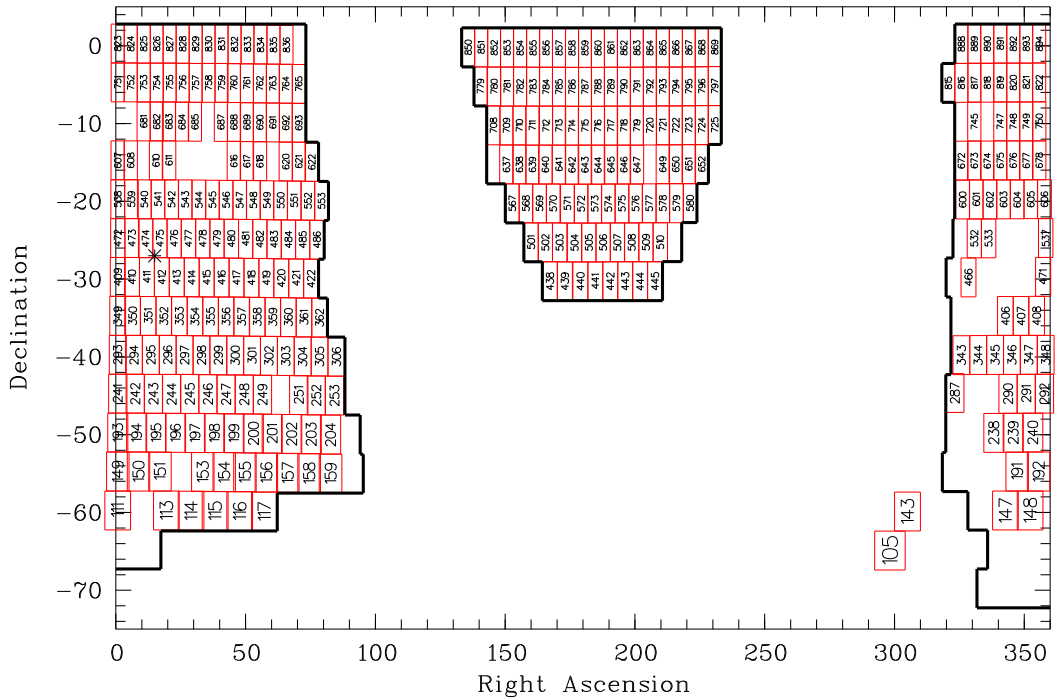, clip=, width=14cm, bbllx=48, bblly=420,
      bburx=527, bbury=740}
    \caption{Definition of HES area (framed) and numbers of fields in which the
      exploitation of the stellar content of the HES is currently carried out.
      For orientation, the position of the southern galactic pole is marked
      with `$\ast$' (lower left corner of field 475). }
  \end{center}
\end{figure*}

The aim of the DA selection in the HES described in this paper is to test the
double-degenerate (DD) scenario for SN~Ia progenitors, in which a binary
consisting of two WDs of large enough mass, merges and produces a
thermonuclear explosion. If this scenario is correct, SN~Ia progenitor systems
should be present among DDs, the latter being identifiable by radial velocity
(RV) variations. Although several double degenerates have been found,
none of them is sufficiently massive to qualify as a viable SN Ia progenitor
\citep[see][]{Maxted/Marsh:1999}. Note however that recently a system
consisting of a sdB and a massive white dwarf has been found, the total mass
of which exceeds the Chandrasekhar mass \citep{Maxtedetal:2000a}. In order to
increase the sample of DDs, a \emph{Large Programme} was proposed to (and
accepted by) the European Southern Observatory (ESO), aiming at observing a
large ($\sim 1\,500$) sample of WDs with VLT Unit Telescope 2 (UT2), and its
high resolution spectrograph UVES, at randomly chosen epochs, dictated by
observing conditions. That is, every time the weather is \emph{too bad} to
carry out Service Mode observations for other programs, WDs are observed.

A program like this requires a large catalog of targets spread all over the
accessible sky in order to be successfull. From the ``Catalog of
Spectroscopically Identified White Dwarfs'' of \cite{McCook/Sion:1999} it is
evident that the southern sky has not yet been surveyed as extensively for
white dwarfs as the northern sky: only 33\,\% of the objects listed are
located at $\delta<0^{\circ}$. Therefore, we were aiming at selecting
additional targets in the data base of the HES.

\begin{figure*}[htbp]
  \begin{center}
    \leavevmode
    \epsfig{file=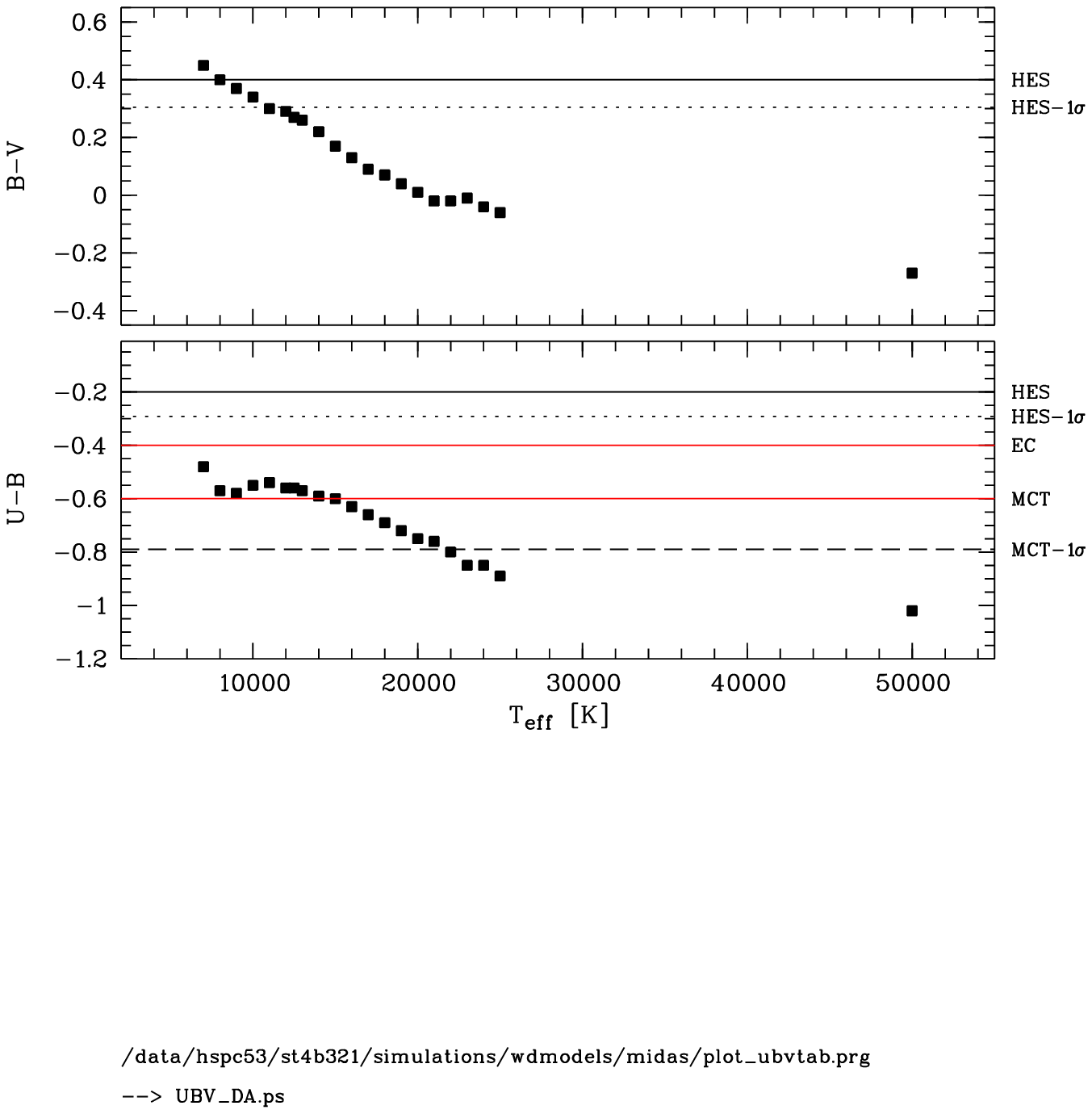, clip=, width=14cm,
      bbllx=67, bblly=250, bburx=460, bbury=514}
  \caption{
    \label{UBV_DA} $U-B$ and $B-V$ for DAs as a function of effective temperature,
    determined with model spectra. Taking into account the r.m.s. error in
    $U-B$ of 0.18\,mag in the MCT survey, the adopted selection criterion
    of $U-B<-0.6$ leads to rejection of cool ($T_{\mbox{\scriptsize eff}}
    \lesssim 22\,000$\,K) DA white dwarfs. In the HES, only DAs of
    $T_{\mbox{\scriptsize eff}}\lesssim 11\,000$\,K are expected to be lost
    when the selection criteria described in Sect. \ref{Sect:DAsel} are
    applied.}  
  \end{center}
\end{figure*}

White dwarfs have been selected from wide angle surveys in the southern
hemisphere before, and also in the HES (see below). ``Classical'' UV excess
surveys, like the Montreal-Cambridge-Tololo survey \citep[MCT;
][]{Demersetal:1986,Lamontagneetal:2000}, or the Edinburgh-Cape survey
\citep[EC; ][]{Stobieetal:1997,Kilkennyetal:1997} can efficiently select
complete samples of hot stars, including WDs. However, completeness at the
\emph{cool} end is either sacrificed for efficiency, as in the MCT (see Fig.
\ref{UBV_DA}), where only objects with $U-B<-0.6$ enter the sample of stars
for which follow-up spectroscopy is obtained \citep{Lamontagneetal:2000}, or
efficiency is sacrificed for completeness, as in the EC. It has been shown
that the EC is 94\,\% complete for objects of $U-B<-0.4$ down to $B=16.5$
\citep{Stobieetal:1997}. However, an intermediate selection step based on
photoelectric $UBV$ photometry has to be used to eliminate the large fraction
($\sim 30$\,\%; see \citealt{Kilkennyetal:1997}) of ``normal'' F and G type
stars.

In the HES, WDs enter the quasar candidate sample if they have $U-B<-0.18$
\citep{hespaperIII}. However, HES quasar candidates are inspected manually at
the computer screen, and in this process hot stars, and stars clearly
exhibiting stellar absorption lines (like e.g. DA white dwarfs, having strong,
broad lines over a wide temperature range; see Fig. \ref{WDmodels} in Appendix
\ref{Sect:SpectralAtlas}) are rejected, and follow-up spectroscopy is
not obtained for them in the course of the quasar survey. This results
in a very efficient quasar selection: typically 70\,\% of the objects for
which follow-up spectroscopy is obtained \emph{are} quasars
\citep{hespaperIII}. The remaining 30\,\% are mainly hot subdwarfs, cool
($T_{\mbox{\scriptsize eff}}\lesssim 20\,000$\,K), helium-rich WDs \citep[DBs,
DZs; see][]{Friedrichetal:2000}, a couple of interesting peculiar objects,
e.g., magnetic DBs \citep{magDB1} and magnetic DAs
\citep{Reimersetal:1994,Reimersetal:1996} have also been discovered in this
way.

The selection of white dwarf candidates described in this paper aims at an
\emph{efficient} selection; that is, the contamination of the sample with
other objects (e.g., hot subdwarfs) shall be as clean as possible, in order not
to waste any observing time at the VLT.

% In Sect. \ref{Sect:HESdatabase} we give a brief overview of the HES;
% in Sect. \ref{FeatureDetection} we describe the automatic feature detection
% in the digital HES spectra.

\section{The HES data base}\label{Sect:HESdatabase}

A description of the HES plate material, plate digitisation and data reduction
can be found in \cite{hespaperIII}. In this section we describe some survey
properties that are particularily important for stellar work in more detail
than it was done in \cite{hespaperIII}, and we briefly repeat a general
description of the HES, for better readability.

%
% Magnitude range
%
The HES is based on IIIa-J plates taken with the 1\,m ESO Schmidt telescope
and its 4$^{\circ}$ prism. It covers the magnitude range $13.0 \gtrsim B_J
\gtrsim 17.5$ \citep{hespaperIII}. The magnitude limits depend on plate
quality.  Note that the value given for the faint limit is the completeness
limit for quasar search, which we define as the magnitude corresponding to
average photographic density in the $B_J$ band $>5\sigma$ above the diffuse plate
background, where $\sigma$ is the background noise. The detection limit of the
HES is approximately one magnitude deeper than the completeness limit. For
stellar applications, the survey magnitude range depends on the object type
searched for. E.g., in our search for metal-poor stars, we only use spectra
that are not affected by saturation effects, which typically start to be
noticable at $B_J\sim 14.0$, and we only include spectra with $S/N>10$ (which
roughly corresponds to $B_J\sim 16.4$), because at lower $S/N$ an efficient
selection of metal-poor stars, by means of a weak or absent Ca~K line, is not
feasible anymore.  However, for most other object types, including DAs, we
adopt the $5\sigma$ magnitude limit.

%
% Wavelength range and spectral resolution
%
The atmospheric cutoff at the blue end, and the sharp sensitivity
cutoff of the IIIa-J emulsion (``red edge'') result in a wavelength
range of $3200\,\mbox{\AA} < \lambda < 5300\,\mbox{\AA}$ (see Fig.
\ref{fig:noisedata_demo}). The spectral resolution of the HES is
primarily seeing-limited.  For plates taken during good seeing
conditions, the pixel spacings chosen in the digitization process
result in an under-sampling, so that in these cases the spectral
resolution is also limited by the sampling.

%
% Sky coverage
% 
The definition of the HES survey area makes use of the mean star density and
average column density of neutral hydrogen for each ESO/SERC field
\citep{hespaperIII}. The adopted criteria roughly correspond to galactic
latitudes of $|b|>30^{\circ}$. The declination range covered by the HES is
$+2.5^{\circ}>\delta> -78^{\circ}$. In result, the survey area consists of 380
fields. Between 1989 and 1998, objective-prism plates were taken for all of
these, and the plates were subsequently digitized and reduced at Hamburger
Sternwarte. As one ESO Schmidt plate covers approximately $5\times 5\deg^2$ on
the sky, the nominal survey area is $9\,500$\,deg$^2$, or the total southern
extragalactic sky. Note, however, that the \emph{effective} survey area is
$\sim 25$\,\% lower, mainly because of overlapping spectra
\citep{hespaperIII}.

%
% Overlap detection
%
\subsection{Detection of overlapping spectra}

Overlapping spectra (hereafter shortly called overlaps) are detected
automatically using the direct plate data of the Digitized Sky Survey I
(DSS~I). For each spectrum to be extracted, it is looked for objects in the
dispersion direction on the direct plate. If there is one, the automatic
procedure marks the corresponding spectrum, so that it can later be excluded
from further processing, if this is desired. {\em It is\/} desired for stellar
work, since the feature detection and object selection algorithms would get
confused otherwise, and a lot of ``garbage'' would enter the candidate
samples. The digital HES data base for stellar work consists of $\sim 4$
million extracted, overlap-free spectra with average $S/N>5$ in the $B_J$
band.

\subsection{Photometry}

As described in \cite{hespaperIII}, the calibration of HES $B_J$
magnitudes is done plate by plate with individual photometric
sequences. The $B_J$ band is formally defined by the spectral
sensitivity curve of the Kodak IIIa-J emulsion multiplied with the
filter curve of a Schott GG395 filter. The overall errors of the HES
$B_J$ magnitudes, including zero point errors, are less than $\pm
0.2$\,mag. Note that $B_J$ can be converted to $B$ using the formula
\begin{equation}
  \label{eq:BJtoB}
 B = B_J + 0.28 \cdot (B-V), 
\end{equation}
which is valid for main sequence stars in the colour range $-0.1<(B-V)<1.6$
\citep{Hewettetal:1995}.

\subsection{Wavelength calibration}

A global dispersion relation for all HES plates was determined by using A-type
stars. In HES spectra of these stars the Balmer lines at least up to H$_{10}$
are resolved (see Fig.  \ref{fig:noisedata_demo}), so that a dispersion
relation can be derived by comparing the $x$-positions (scan length in
$\mu$\/m) of these lines with the known wavelengths.  \cite{Borraetal:1987}
used the position of the ``red edge'' of objective-prism spectra to determine
the zero point of the wavelength calibration, but noticed that the position
depends on the energy distribution of the object. Therefore, in the HES we
decided to use a zero point specified by an astrometric transformation between
direct plates and spectral plates. The wavelength calibration is accurate to
$\pm 10\,\mu$\/m. This corresponds to $\pm 4.5$\,{\AA} at H$\gamma$ and $\pm
2.3$\,{\AA} at $\lambda = 3500$\,{\AA}.

\subsection{Estimation of the amplitude of pixel-wise
  noise}\label{sect:NoiseEstimation}

\begin{figure}[htbp]
  \begin{center}
    \leavevmode
    \epsfig{file=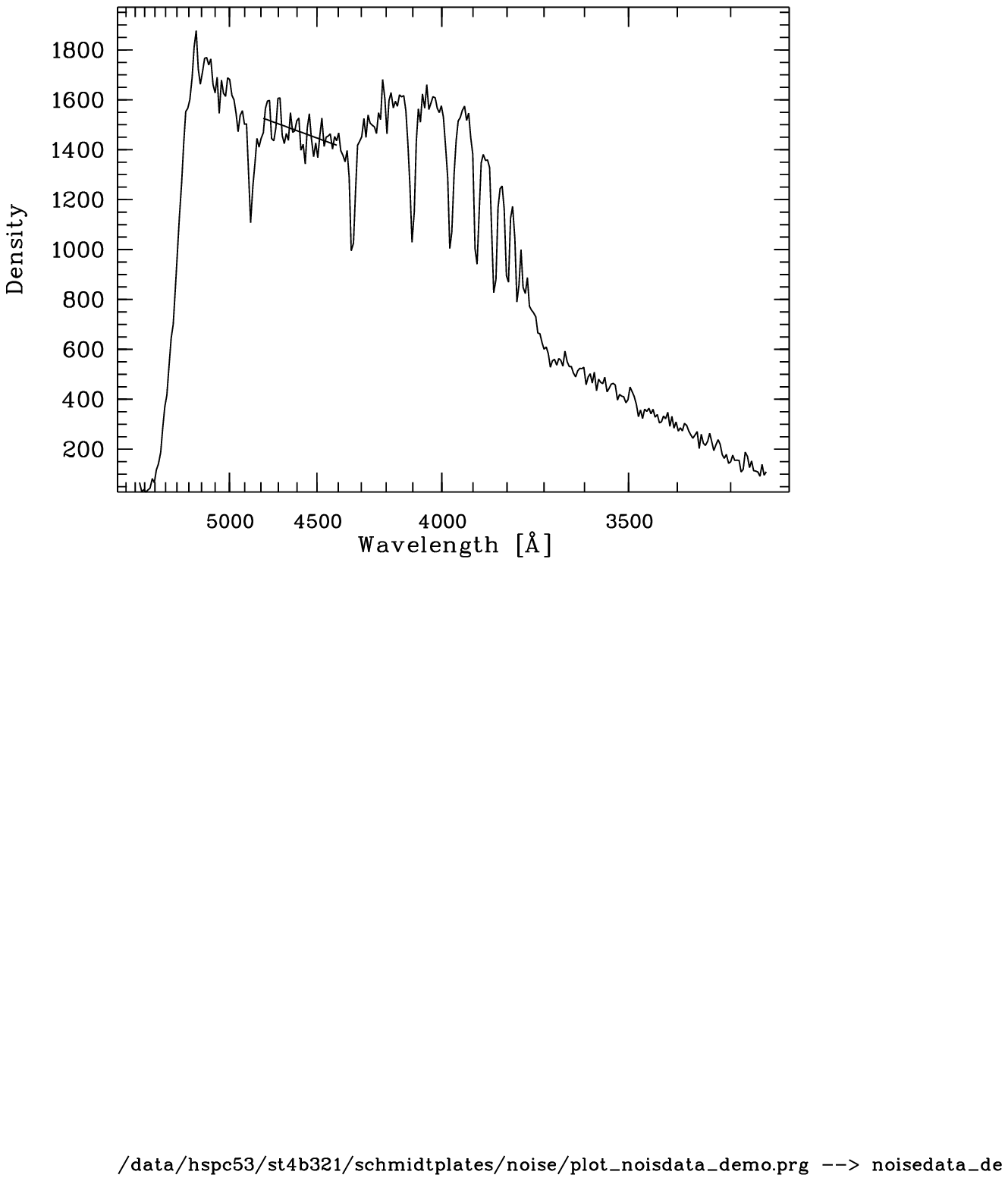, clip=, width=8.8cm,
      bbllx=66, bblly=426, bburx=371, bbury=640}
  \caption{\label{fig:noisedata_demo}
    Measurement of noise in absorption line free spectral region of A-type
    stars. Note that we plot HES objective prism spectra such that wavelengths
    are \emph{decreasing\/} towards the right, because the scan length $x$ is
    \emph{increasing\/} in this direction. The sharp drop of the spectra at
    the red end is due to the IIIa-J emulsion sensitivity cutoff at $\sim
    5400$\,{\AA}.}
  \end{center}
\end{figure}

Following the approach of \cite{Hewettetal:1985}, we determine the amplitude
of pixel-wise noise as a function of photographic density $D$ plate by plate
using A- and F-type stars.  A straight line fit is done to the spectral region
between H$\beta$ and H$\gamma$ (see Fig.  \ref{fig:noisedata_demo}).
The $1\,\sigma$-scatter around this pseudo-continuum fit is taken as noise
amplitude. In this approach we assume that the scatter is mainly due to noise,
since in early-type stars the spectral region under consideration includes
only very few absorption lines at the spectral resolution of the HES.
Moreover, it is expected that the population of A- and F-type stars found at
high galactic latitudes is dominated by metal-poor stars, so that metal lines
are usually very weak. However, we can not exclude that we overestimate the
noise by a few percent due to contributions of metal lines to the scatter
about the pseudo-continuum fit.

We measure the noise amplitude for all A- and F-type stars present on
each HES plate (typically several hundred per plate), and compute the mean
density $D$ in the fit region. This yields data points $(D,\mbox{noise})$, to
which a 2nd order polynomial is fitted, i.e.
\begin{equation}
 \mbox{noise} = a_0 + a_1\cdot D + a_2\cdot D^2. \label{Parabelgleichung}
\end{equation}
\begin{figure}[htbp]
  \begin{center}
    \leavevmode
    \epsfig{file=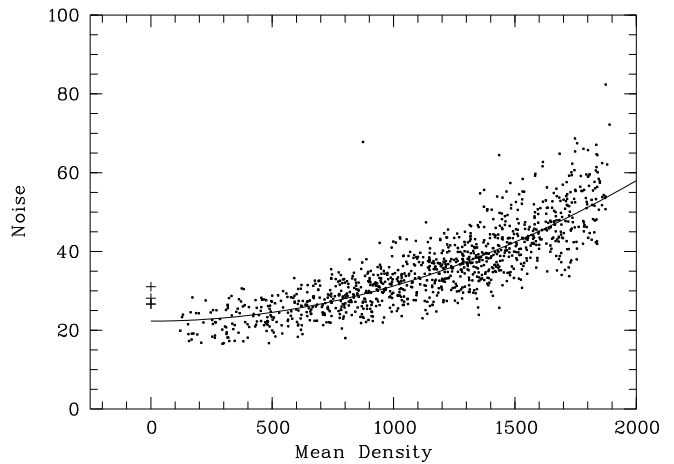, clip=, width=8.8cm,
      bbllx=58, bblly=426, bburx=367, bbury=644}
  \caption{\label{fig:noisefit}
    Fitting of a 2nd order polynomial to data points $(D,\mbox{noise})$,
    for estimation of pixel-wise noise as a function of density. The plate
    background noise of each plate quarter is marked with `$+$'. It is
    always higher than expected from the noise parabola fit, since the
    optimal spectrum extraction algorithms result in a bias towards lower
    noise.}
  \end{center}
\end{figure}
Note that in the HES, $D$ refers to density above
diffuse plate background (bgr) in arbitrary units called counts. The relation
between counts and photographic densities $D_{\mbox{\scriptsize photo}}$ is
\begin{equation}\label{Dphoto}
  D_{\mbox{\scriptsize photo}} = \frac{D\,\mbox{[counts]}
    + D_{\mbox{\scriptsize bgr}}\,\mbox{[counts]}}{800}.
\end{equation}

\begin{figure*}[htbp]
  \begin{center}
    \leavevmode
    \epsfig{file=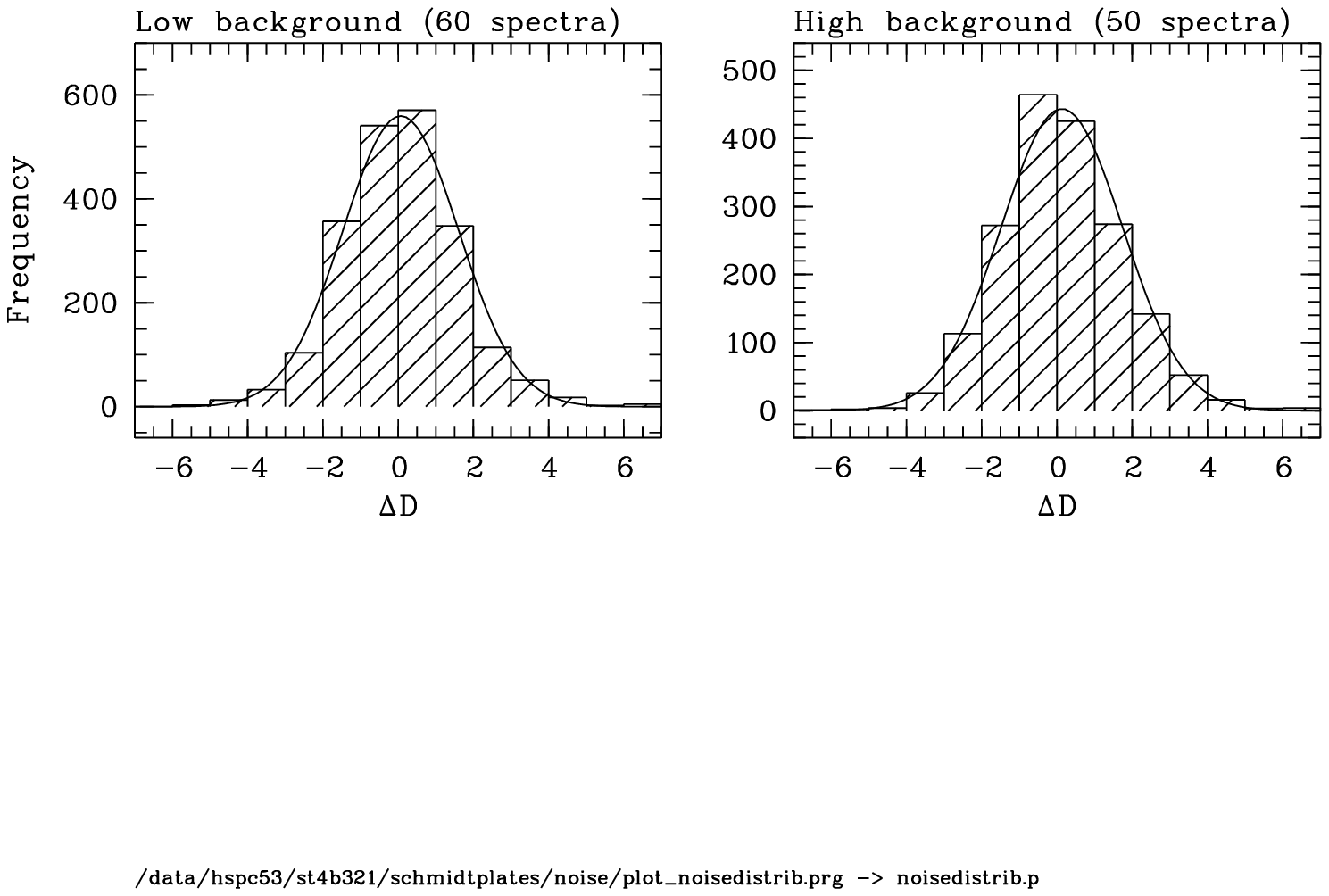, clip=, width=14cm,
      bbllx=69, bblly=566, bburx=499, bbury=736}
  \caption{\label{noisedistrib}
    Distribution of pixel-wise noise in absorption line free regions of
    60 spectra of A-type stars from seven plates with low sky background (left
    panel), and 50 spectra from eight plates with high background (right
    panel). For further explanation see text.}  
  \end{center}
\end{figure*}

\noindent For the determination of the coefficients $a_0\dots a_2$ we use a robust fit
algorithm which minimizes the sum of absolute deviations, taking into account
the following set of boundary conditions:
\begin{description}
\item[$a_0 > 0$.] The noise at $D=0$ is the noise of the plate background,
  which is always $>0$.
\item[$a_2\ge 0$.] Since $D \ge 0$, and the noise increases monotonically with
  $D$, $a_2$ must be positive (or zero).
\item[$a_1\ge 0$.] From the previous boundary condition follows that the
  polynomial has a minimum at $D_{\mbox{\scriptsize min}}=-a_1/(2a_2)$. Using
  again the argument that the noise increases monotonically with $D$, it
  follows that $D_{\mbox{\scriptsize min}}\le 0$. Since $a_2\ge 0$, $a_1$ must
  be $\ge 0$.
\end{description}
An example for a such a fit is shown in Fig. \ref{fig:noisefit}.

\label{par:noisdistrib} For simulations of spectra (see
Sect. \ref{slit2objprism}) it is very important to know the \emph{form} of the
distribution of noise. We investigated this by using 50 spectra of A-type
stars from eight plates with high sky background ($D_{\mbox{\scriptsize
bgr}}>1\,500$), and 60 spectra from seven plates with low background
($D_{\mbox{\scriptsize bgr}}<700$). These spectra were chosen by hand from the
sample of automatically selected A-type stars, in order to ensure that
misclassified spectra, and spectra for which the fit of the continuum between
H$\beta$ and H$\gamma$ by a straight line is not fully adequate, do not
confuse the results. Five of the original set of 115 spectra were excluded in
the manual selection process.

The deviations from the continuum fits were collected for each spectrum,
shifted to a median of zero, and divided by the average of the absolute values of
the upper and lower $50$\,\% quartile, so that a comparison of the noise
distributions measured in different spectra (with different noise amplitude)
is possible. The result is that the distribution of pixel-wise noise is
almost perfectly Gaussian, independently of plate background (see Fig.
\ref{noisedistrib}).

\begin{figure}[htbp]
  \begin{center}
    \leavevmode
    \epsfig{file=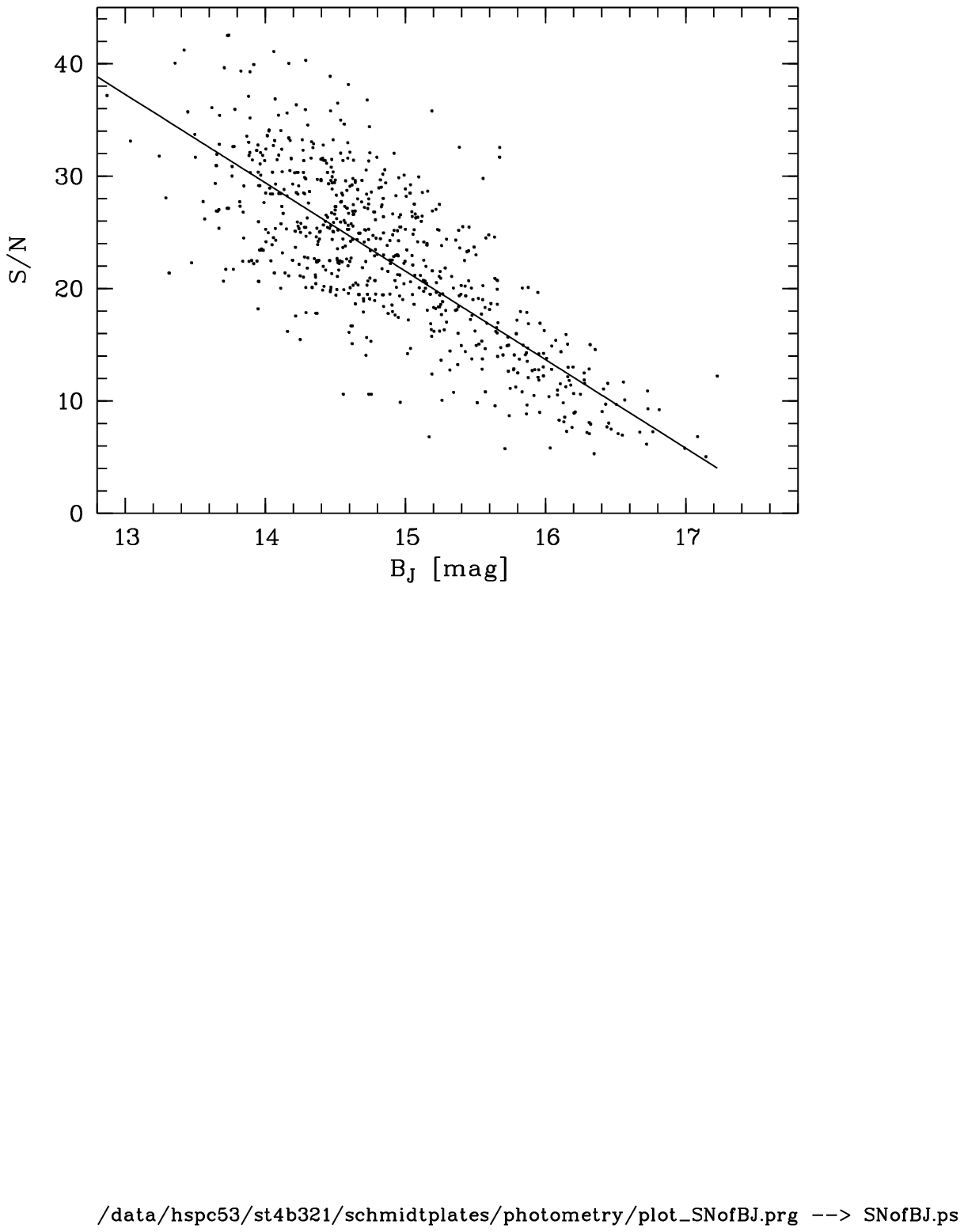, clip=, width=8.8cm,
      bbllx=63, bblly=426, bburx=357, bbury=640}
  \caption{\label{SNofBJ} Average pixel-wise $S/N$ in the $B_J$ band as a
    function of $B_J$ for 589 not saturated point sources from more than 100
    different HES plates. The straight line fit corresponds to
    $B_J=17.7-0.13\cdot S/N$.}  
  \end{center}
\end{figure}

Fig. \ref{SNofBJ} shows the relation between average pixel-wise $S/N$ in the
$B_J$ band and $B_J$ magnitudes for 589 not saturated point sources from many
different HES plates.  The large scatter of $S/N$ for given $B_J$ shown in
this plot is due to varying plate background, and seeing.

\section{Feature detection}\label{FeatureDetection}

\subsection{Detection of stellar lines}\label{sect:StellarLineDetection}

It is critical for all quantitative selection methods to have a set of
\emph{reliable} features at hand. The total set of available features should
contain as much information of the objects to be classified as possible.
Therefore, we implemented a flexible, robust algorithm which allows to detect
stellar absorption and/or emission lines in HES spectra.
\cite{Hewettetal:1985} used a template matching technique to automatically
detect absorption and emission lines in objective-prism spectra. Tests with
HES spectra have shown that a fit algorithm involving a couple of boundary
conditions (see below) is much more stable and leads to more accurate
measurements of equivalent widths than template matching. The algorithm used
in the HES consists of the following steps:

\begin{enumerate}
\item[1.] Determination of continuum by filtering with a wide median filter
  and narrow Gaussian filter. A similar filtering technique was also
    used by \cite{Hewettetal:1985} and \cite{Borraetal:1987}.
\item[2.] Improvement of determination of the wavelength calibration zero
  point by fitting of 3 sets of stellar lines. The sets contain the strongest
  stellar absorption lines of early type, solar type, and late type stars,
  respectively. The individual line depths, and the zero point offset of
  wavelength calibration are fitted simultaneously. The \emph{relative}
  positions of the stellar lines are held fixed, and the line \emph{widths} is
  held fixed at the value of the seeing profile widths, which is measured
  during spectral extraction. The set of lines giving the strongest signal,
  i.e. largest average equivalent widths, is selected, and the wavelength
  calibration zero point determined with that fit is adopted.  
\item[3.] Improvement of continuum determination:
  \begin{enumerate}
  \item[a.] Fitting of \emph{all} stellar lines detectable in HES spectra
  \item[b.] Subtraction of fitted lines from the original spectrum
  \item[c.] Computation of improved continuum by filtering the line-subtracted
    spectrum again with a wide median filter and narrow Gaussian filter
  \item[d.] Go back to 3a, if $n_{\mbox{\scriptsize iter}}<3$; otherwise
    compute rectified spectrum with final continuum.
  \end{enumerate}
\item[4.] Fitting of all stellar lines in the rectified spectrum by
  Gaussians.
\end{enumerate}
For each spectral line it can be chosen whether it is to be detected in
absorption or emission. The output of the fit algorithm are equivalent width,
FWHM and $S/N$ of the lines, and shift of the wavelength calibration zero
point. Any spectral lines not yet considered can easily be included by just
adding its wavelength to the list of lines to be fitted. In this work we
make use of the equivalent width of H$\beta$, H$\gamma$ and H$\delta$ only,
which are summed to the parameter {\tt balmsum}.

\subsection{Broad band and intermediate band colours}

\begin{figure}[htbp]
  \begin{center}
    \leavevmode
    \epsfig{file=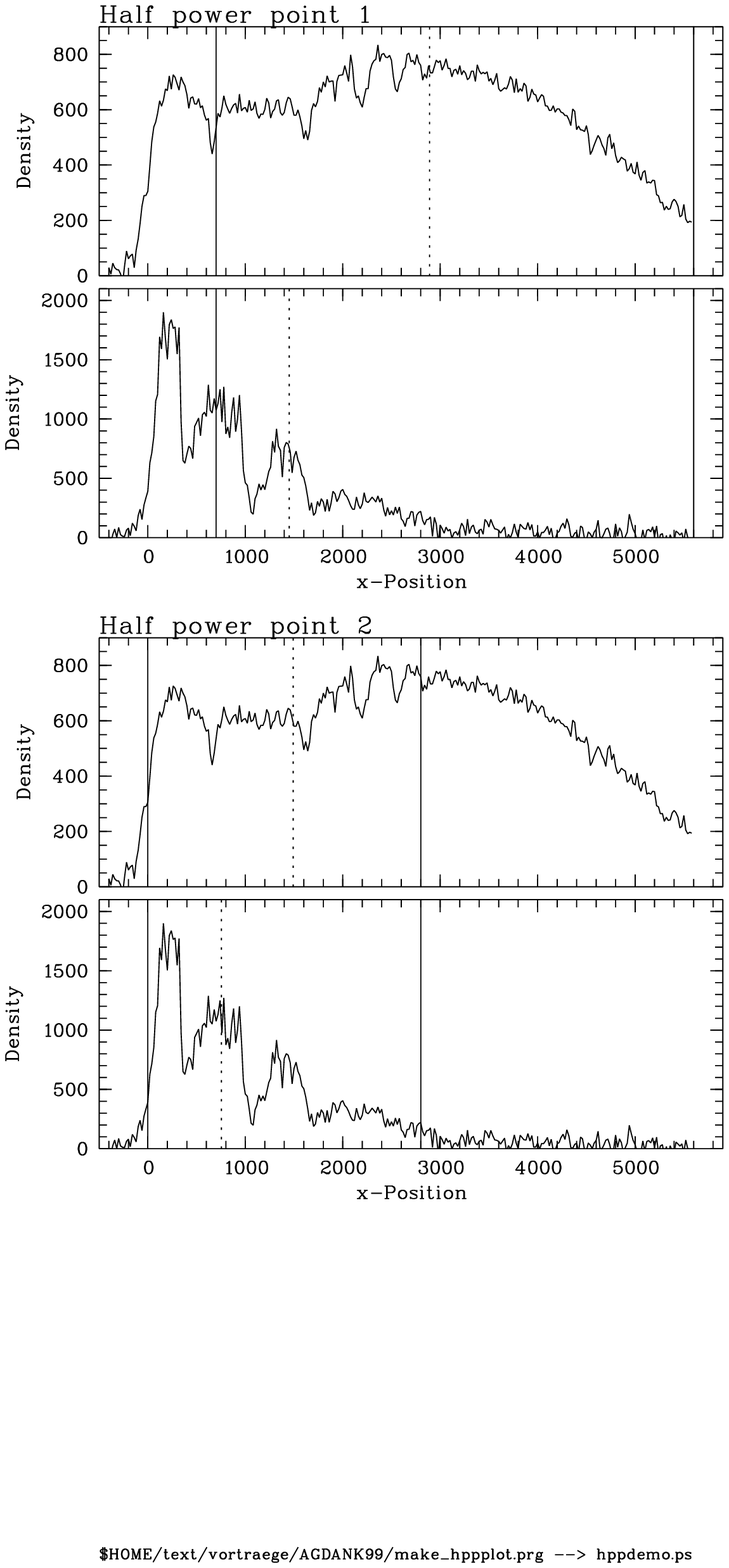, clip=, width=8.8cm,
      bbllx=95, bblly=200, bburx=428, bbury=748}
  \caption{\label{hppdemo}
    Illustration of spectral half power points \texttt{x\_hpp1} and
    \texttt{x\_hpp2}. Solid lines mark the regions in which the hpps
    are computed; dotted lines indicate the position of the hpps. }  
  \end{center}
\end{figure}

As already noticed by \cite{Hewettetal:1985}, relative colours can be
determined quite accurately directly from objective-prism spectra.
\cite{Hewettetal:1995} used so-called ``half power points (hpps)'' to measure
relative colours, that is, bisecting points of the photographic density
distribution. hpps are equivalent to broad-band colours, but have the
advantage of being more robust against noise. \cite{hespaperIII} introduced
half power points that are computed only for a \emph{part} of the spectrum
(see Fig. \ref{hppdemo}). The half power points \texttt{x\_hpp1} and
\texttt{x\_hpp2} are well correlated with $U-B$ and $B-V$, respectively.
  
Since it is helpful in many stellar applications to have not only relative,
but calibrated $U-B$ and $B-V$ colours at hand, we established calibrations of
\texttt{x\_hpp1} and \texttt{x\_hpp2} versus $U-B$ and $B-V$, respectively.

\begin{figure}[htbp]
  \begin{center}
      \epsfig{file=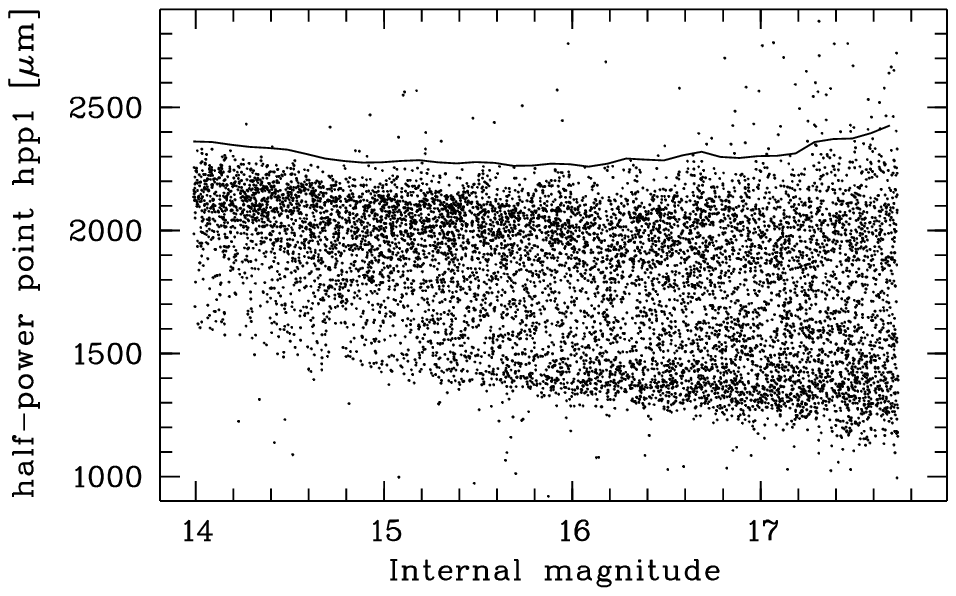, clip=, width=8.8cm,
        bbllx=68, bblly=86, bburx=342, bbury=258}
  \caption{\label{hpp1_cutoff} Cutoff-line for bisecting point \texttt{x\_hpp1} 
     on one HES plate.}
  \end{center}  
\end{figure}

A more precise colour calibration can be achieved when distances \texttt{dx}
to a cutoff line in a colour-magnitude diagram (see Fig. \ref{hpp1_cutoff}) is
used instead of \texttt{x} values (scan length in $\mu$\/m) for the bisecting
point, because plate-to-plate variations of the spectral sensitivity curves
are compensated in this way. The cutoff line separates the bulk of ``normal''
stars from UV-excess objects (or objects with unusually low $B-V$ in case of
\texttt{dx\_hpp2}). The cutoff is determined by a break finding algorithm.

\begin{figure*}[htbp]
  \begin{center}
      \epsfig{file=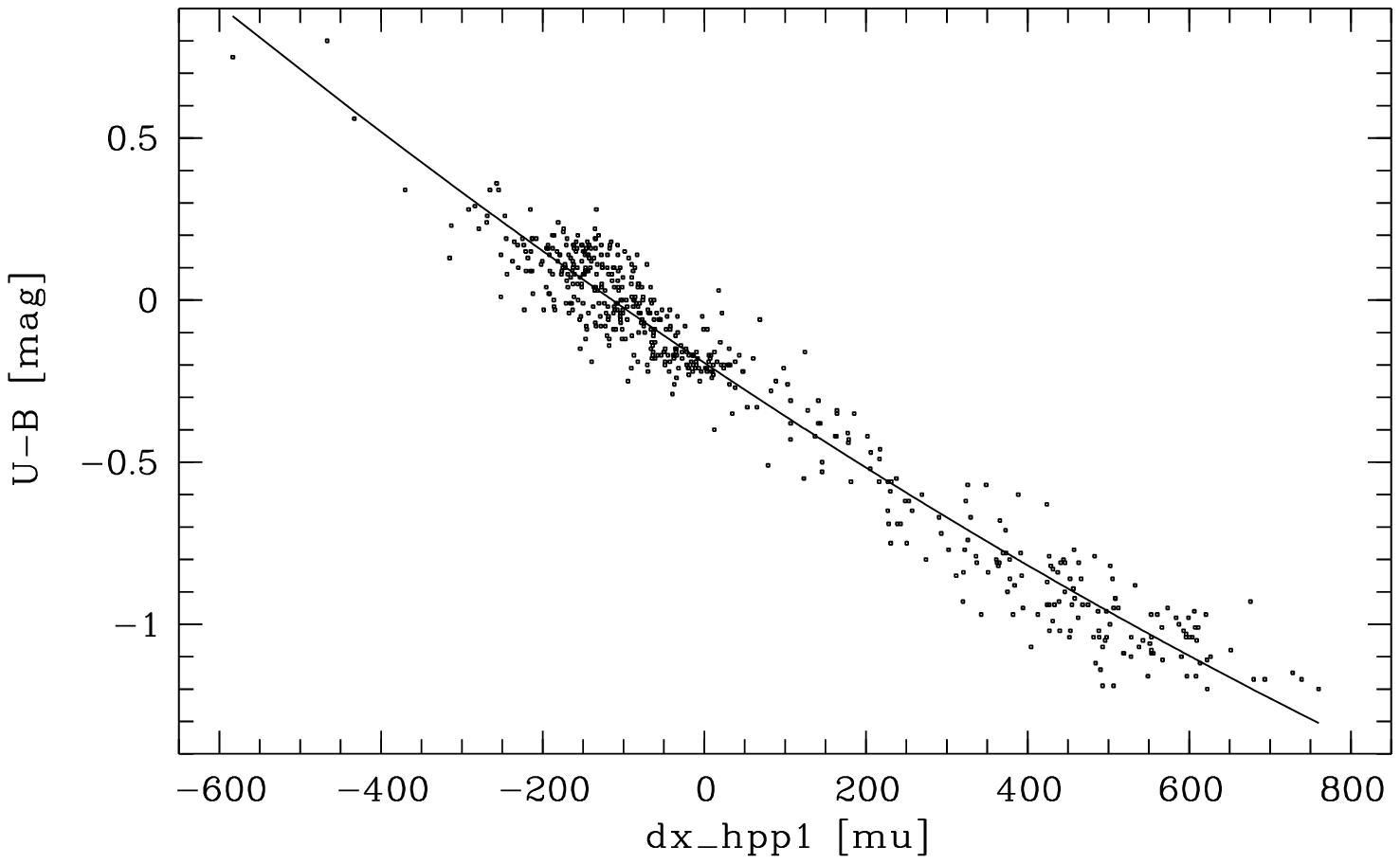, clip=, width=8cm, height=5.558cm,
        bbllx=43, bblly=420, bburx=470, bbury=683}\hspace{0.3cm}
      \epsfig{file=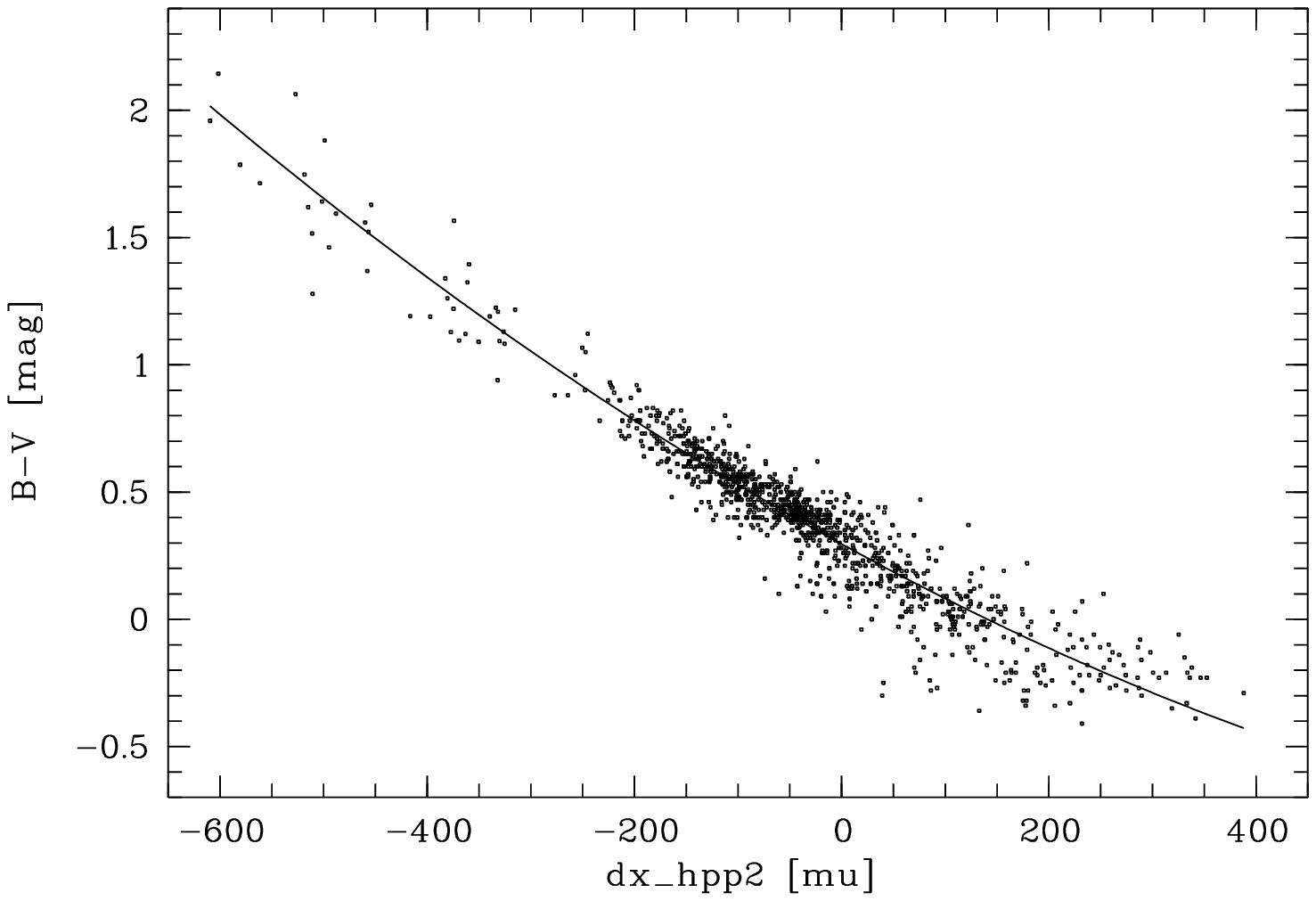, clip=, width=8cm, height=5.5cm,
        bbllx=42, bblly=420, bburx=471, bbury=712}
  \caption{\label{dxhppfit} Calibration of \texttt{dx\_hpp1} versus $U-B$
    using a sample of 573 objects from the EC and HK surveys present on HES
    plates (left panel); calibration of \texttt{dx\_hpp2} versus $B-V$
    using spectra of 1256 objects (right panel).}
  \end{center}
\end{figure*}

Because the blue end of the HES spectra is sensitive to contamination by
overlaps, special care must be taken to exclude such spectra from the
calibration of \texttt{dx\_hpp1}. This was done by applying stricter overlap
selection criteria. In addition, an iterative $\kappa\sigma$-clipping with
$\kappa=3$ was employed to exclude overlaps unrecognized by the automatic
detection. 50 of the 623 spectra in the original data set were rejected, so
that the calibration uses spectra of 573 objects (see Fig. \ref{dxhppfit}; a
similar plot was shown in \citealt{hespaperIII}).

A potential problem for the $B-V$ calibration is that the $V$ band is not
fully covered by the HES wavelength range. Therefore, the calibration for very
red objects is inaccurate, or even impossible. As calibrators for red objects,
36 carbon stars were used, for which $BV$ photometry was obtained 
at the ESO 2.2\,m telescope in April 1999. Carbon stars with $B-V>2.5$
were excluded from the fit. For $B-V\lesssim 1.0$, 778 stars from the HK
survey of \cite{BPSII}, 354 FHB and other A-type stars of
\cite{Wilhelmetal:1999b}, and 272 objects from the northern galactic cap
fields of the EC survey \citep{Kilkennyetal:1997} present on HES plates were
used.  Linear fits in three colour regions were done separately, in
order to evaluate the scatter independently, and check consistency. Then, a
combined fit to all 1256 unique objects was done (see Fig. \ref{dxhppfit}).

The results of the fits are summarized in Tab. \ref{ColourFits}. Note that a
single fit contains objects from a large fraction of the 329 stellar HES
plates, and a wide range of object types, e.g. carbon stars, metal-poor stars,
solar metallicity F- and G-type stars, field horizontal branch A-type stars,
``normal'' A-type stars, DAs, DBs, sdBs and quasars. The achieved accuracies
are $\sigma_{U-B}=0.092$\,mag, and $\sigma_{B-V}=0.095$\,mag for the $B-V$ fit
using all calibration objects together.  The accuracy in $B-V$ for red
($B-V\gtrsim 1$) and blue ($B-V\lesssim 0.3$) objects is a factor of $\sim 2$
worse ($\sigma=0.15$\,mag and $0.12$\,mag, respectively) than for intermediate
$B-V$ objects ($\sigma=0.074$\,mag).

\begin{table}[htbp]
  \caption{\label{ColourFits} Broad- and intermediate-band colour calibration fits.}
  \begin{flushleft}
    \begin{tabular}{lcrl}\hline\hline
      Colour & \multicolumn{1}{l}{valid range} &
      N$_{\mbox{\scriptsize stars}}$ & $\sigma$ [mag] \rule{0.0ex}{2.3ex}\\[0.2ex]\hline
%-----------------------------------------------------------------------%
      $B-V$ & $-0.6<B-V<2.0$ & $1259$ & $0.095$\rule{0.0ex}{2.3ex}\\
%-----------------------------------------------------------------------%
      $U-B$ & $-1.4<U-B<0.8$ & $573$ & $0.092$\\
%-----------------------------------------------------------------------%
      $c_1$ & $-0.4<c_1<1.0$ & $79$  & $0.15$\\\hline\hline
%-----------------------------------------------------------------------%
    \end{tabular}
  \end{flushleft}
\end{table}

\begin{figure}[htbp]
  \begin{center}
    \leavevmode
    \epsfig{file=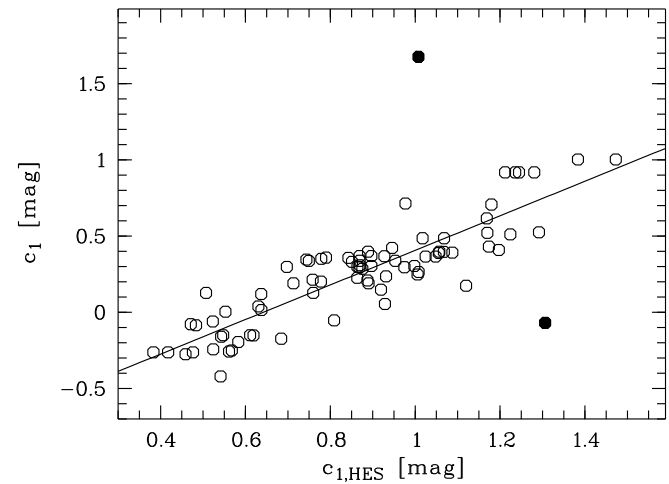, clip=, width=8.8cm,
      bbllx=54, bblly=425, bburx=328, bbury=626}
  \caption{\label{fig:c1calib} Calibration of Str\"omgren $c_1$ measured in
    HES spectra. The two filled circles mark objects that were excluded from
    the fit.}  
  \end{center}
\end{figure}

We obtain Str\"omgren coefficients $c_1=(u-v)-(v-b)$ directly from HES spectra
by averaging the density in the Str\"omgren $uvb$ bands, and computing
internal coefficients $c_{1,\mbox{\tiny HES}}$ from that. $c_{1,\mbox{\tiny
    HES}}$ was calibrated using a total of 79 stars not saturated in the HES,
from three different sources. 22 metal-poor stars were taken from
\cite{Schusteretal:1996}, 43 stars from Beers (2000, priv.  comm.), of which 2
were rejected as outliers (see Fig.  \ref{fig:c1calib}), and 16 hot subdwarfs
from an updated version of the catalog of \cite{Kilkennyetal:1988}.  The
$1\,\sigma$ error of the calibration is $0.15$\,mag.  $c_1$ can be used as a
gravity indicator for early-type stars, since it measures the strength of the
Balmer discontinuity.

\section{Simulation of objective-prism spectra}\label{slit2objprism}

In many stellar applications of the HES, it is not possible to generate large
enough training and test samples from \emph{real} spectra present on HES
plates. This is because usually the target objects are very rare. Therefore,
we have developed methods to generate \emph{artificial} learning samples by
simulations, using either model spectra, or slit spectra. In this paper
we will use the simulations for the development of selection criteria.
In later papers we will use sets of simulated spectra as learning samples for
selection of e.g. metal-poor stars by automatic spectral classification.

The conversion of model spectra, or slit spectra, to objective-prism spectra
consists of five steps:
\begin{enumerate}
\item[(1)] Rebinning to the non-equidistant pixel size according to the
  global dispersion relation for the HES
\item[(2)] Multiplication with HES spectral sensitivity curve(s)
\item[(3)] Smoothing with a Gaussian filter, for simulation of seeing
\item[(4)] Adding of pixel-wise, normally distributed noise
\item[(5)] Random shift of the simulated spectrum according to the error
  distribution of the wavelength calibration zero point ($\pm 10\,\mu$\/m).
\end{enumerate}
Step (4) ensures that objects of any brightness can be simulated. The 
$B_J$ magnitude range corresponding to a given $S/N$ can be read from
Fig. \ref{SNofBJ}.

\subsection{HES spectral sensitivity curves}

Spectral sensitivity curves (SSCs) for HES plates were determined by
comparison of WD \emph{model} spectra, rebinned to the wavelength dependent
pixel size $\Delta\lambda$ of the objective-prism spectra, with
objective-prism spectra of DAs on HES plates.  We do not use the slit spectra
directly as reference, because slit losses would produce erroneous results.
The theoretical DA spectra, taken from a standard grid of WD model atmospheres
\cite[see][for a description]{Finleyetal:1997}, were fitted to the slit
spectra with methods described in \cite{Finleyetal:1997} and
\cite{Homeieretal:1998}. The spectra were taken with the Boller \& Chivens
spectrograph attached to the ESO 1.52\,m telescope, and with DFOSC at the
ESO-Danish 1.54\,m telescope. The atmospheric parameters of the twelve objects
used in this investigation are listed in Tab.  \ref{tab:ScurveBase}

\begin{figure}[htbp]
  \begin{center}
    \leavevmode
    \epsfig{file=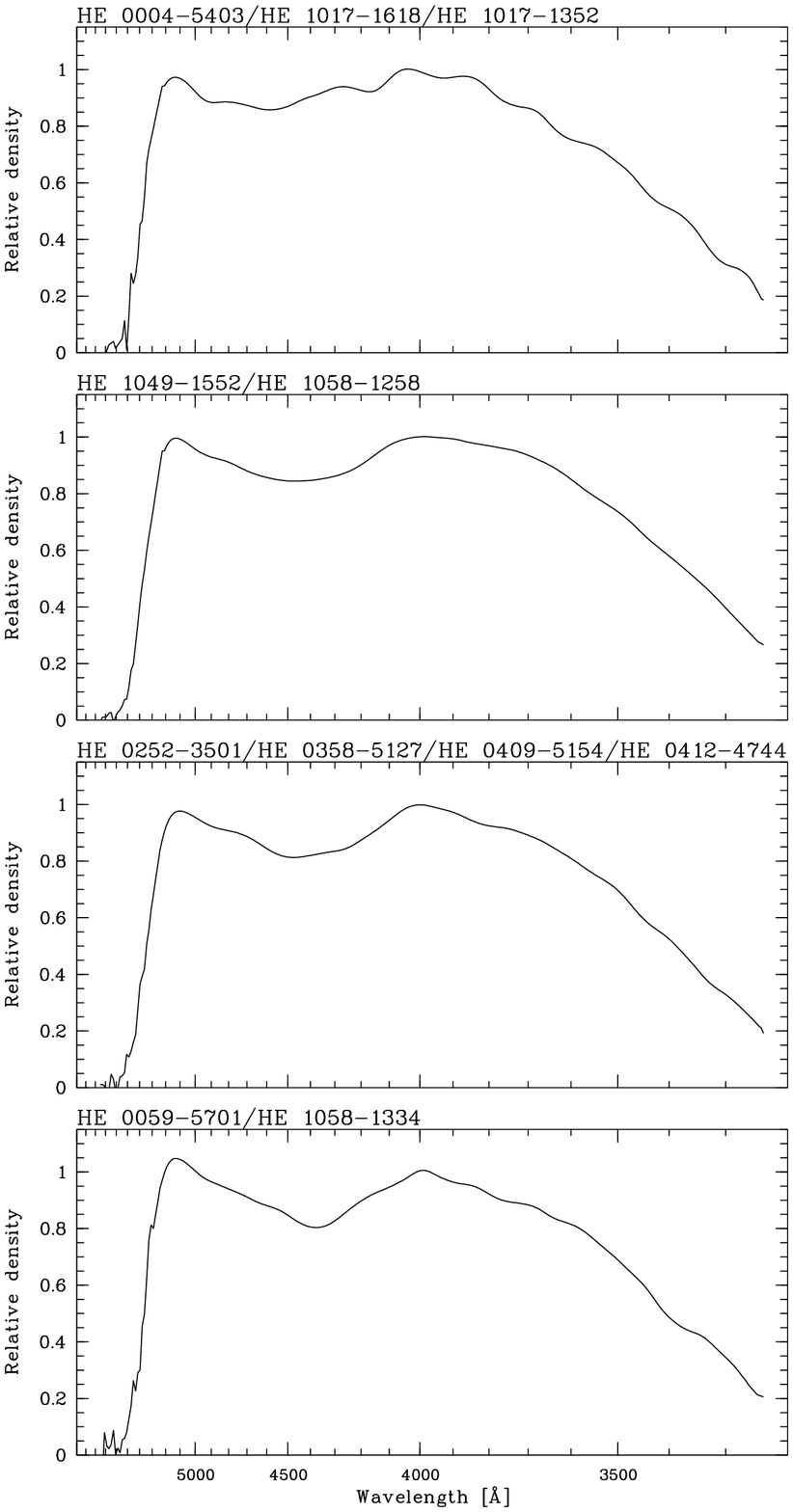, clip=, width=8.8cm,
      bbllx=103, bblly=44, bburx=485, bbury=766}
  \caption{\label{scurvecat} Averaged spectral sensitivity curves. }  
  \end{center}
\end{figure}

By comparing SSCs for plates from different plate batches, with different sky
background, and generated with objects spanning a wide brightness range (see
Tab. \ref{tab:SummarizedScurves}), but below the saturation threshold, we
investigated the possible systematic influence of these characteristics on the
shape of the SSCs.  By comparing the shapes of the twelve resulting SSCs, we
found that there is \emph{no} systematic influence of object brightness, plate
batch and sky background on SSC shape. The plate material of the HES is
surprisingly homogenous; however, a \emph{slight} variation of SSC shape is
present, which hence must be attributed to another parameter.  Since it is the
blue part of the SSCs that varies, it is very likely that the time span
between hypersensitization and development of the plate is responsible for the
shape variations.

\begin{figure*}[htbp]
  \begin{center}
    \leavevmode
    \epsfig{file=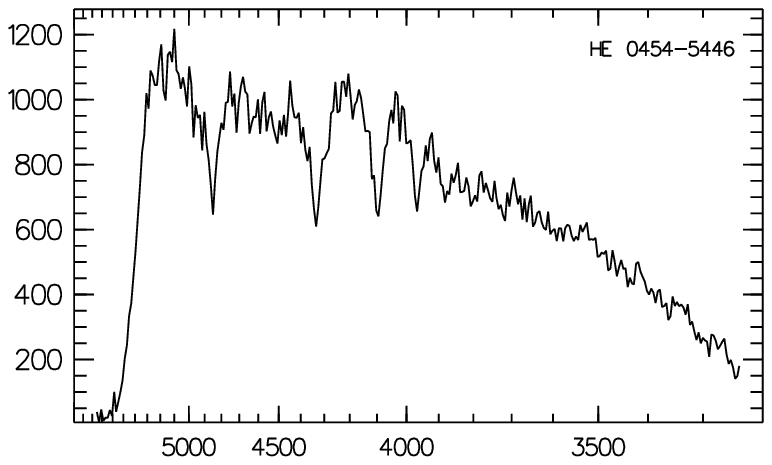, clip=,  width=7.5cm, 
      bbllx=90, bblly=638, bburx=310, bbury=771}\hspace{0.5cm}
    \epsfig{file=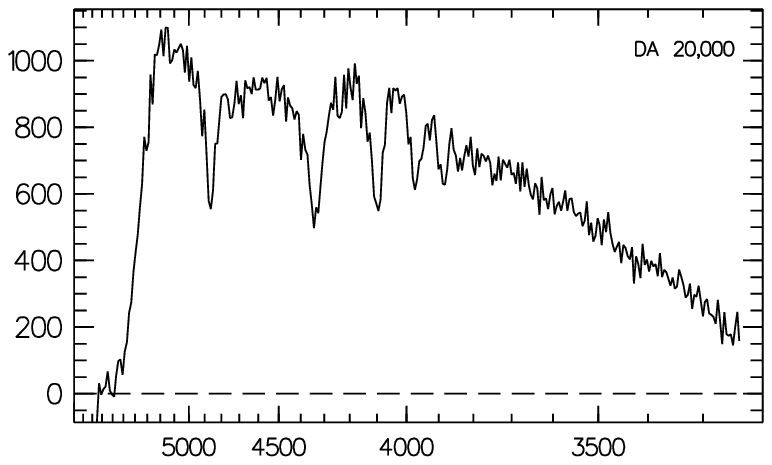, clip=,  width=7.5cm, 
      bbllx=90, bblly=638, bburx=310, bbury=771}
  \caption{\label{slit2objprismdemo} HES spectrum of HE\,0454--5446, a white
    dwarf of type DA (left panel), in comparison with a model spectrum of a
    DA with $T_{\mbox{\tiny eff}}=20\,000$\,K, converted into an objective-prism
    spectrum and with added artificial noise (right panel). The units of the
    ordinates are photographic densities in arbitrary units.}
  \end{center}
\end{figure*}

\begin{table}[htbp]
  \caption{\label{tab:ScurveBase} Atmospheric parameters of the DA white dwarfs 
    used for determination of spectral sensitivity curves. In the last column
    we give additional identifiers for objects listed in \cite{McCook/Sion:1999}.}
  \begin{flushleft}
    \begin{tabular}{lccl}\hline\hline
      Name & $T_{\mbox{\scriptsize eff}}$ [K]& $\log g$ & McCook \& Sion\rule{0.0ex}{2.3ex}\\[0.2ex]\hline
      HE 0004--5403 & $18\,200\pm 300$ & $8.26\pm 0.06$ & \rule{0.0ex}{2.3ex}\\
      HE 0059--5701 & $30\,400\pm 300$ & $8.08\pm 0.06$ & \\
      HE 0252--3501 & $17\,400\pm 300$ & $7.35\pm 0.05$ & WD 0252--350\\
      HE 0358--5127 & $24\,100\pm 300$ & $8.10\pm 0.05$ & \\
      HE 0409--5154 & $27\,500\pm 300$ & $8.00\pm 0.06$ & \\
      HE 0412--4744 & $19\,300\pm 300$ & $8.08\pm 0.06$ & \\
      HE 0418--5326 & $27\,900\pm 200$ & $8.00\pm 0.05$ & \\
      HE 1049--1552 & $20\,200\pm 200$ & $8.63\pm 0.04$ & WD 1049--158\\
      HE 1058--1258 & $24\,700\pm 200$ & $8.84\pm 0.04$ & WD 1058--129\\
      HE 1058--1334 & $15\,900\pm 300$ & $8.00\pm 0.07$ & \\
      HE 1017--1618 & $28\,600\pm 300$ & $8.30\pm 0.06$ & WD 1017--163\\
      HE 1017--1352 & $33\,500\pm 200$ & $8.25\pm 0.05$ & \\\hline\hline
    \end{tabular}
  \end{flushleft}
\end{table}

We grouped the twelve SSCs into four SSC classes of similar shape, and
averaged them within these classes (see Fig. \ref{scurvecat}). When converting
model spectra or slit spectra to objective prism spectra, we use an SSC
created by averaging the four averaged SSCs with randomly assigned weights.

\begin{table}[htbp]
  \caption{\label{tab:SummarizedScurves} Averaging of spectral sensitivity
    curves of similar shape.  {\tt bgr} is the diffuse background (in counts) 
    averaged over four plate quarters.}
  \begin{flushleft}
    \begin{tabular}{llcrrl}\hline\hline
      \# & Name & $B_J$ & \multicolumn{1}{c}{Plate} & {\tt bgr} &
      Batch\rule{0.0ex}{2.3ex}\\\hline      
    1 & HE 0004--5403 & 16.2 & 12076 & 1123 & 1D4\rule{0.0ex}{2.3ex}\\
    1 & HE 1017--1618 & 15.8 &  8402 & 1363 & 1K6\\
    1 & HE 1017--1352 & 14.4 &  8402 & 1363 & 1K6\\\hline
    2 & HE 1049--1552 & 14.2 &  9091 &  752 & 1C8\rule{0.0ex}{2.3ex}\\
    2 & HE 1058--1258 & 14.8 &  9091 &  752 & 1C8\\\hline
    3 & HE 0252--3501 & 16.0 & 11420 & 1039 & 1D4\rule{0.0ex}{2.3ex}\\
    3 & HE 0358--5127 & 15.4 & 10844 &  765 & 1I3\\
    3 & HE 0409--5154 & 16.1 & 10844 &  765 & 1I3\\
    3 & HE 0412--4744 & 16.5 & 10844 &  765 & 1I3\\\hline
    4 & HE 0059--5701 & 16.4 & 12052 & 1026 & 1D4\rule{0.0ex}{2.3ex}\\
    4 & HE 1058--1334 & 16.6 &  9091 &  752 & 1C8\\\hline\hline
    \end{tabular}
  \end{flushleft}
\end{table}

\subsection{Adding noise}

We add artificial, normally distributed noise to the converted spectra, in
order to simulate objective-prism spectra of a desired brightness. We parameterize
the $S/N$ of a spectrum by the mean $S/N$ in the $B_J$ band,
\begin{displaymath}
  \overline{\left(\frac{S}{N}\right)}_{B_J}
  =\frac{1}{n}\sum_{i=1}^{n}\frac{D_i}{a_0+a_1D_i+a_2D_i^2},
\end{displaymath}
using the noise model described in Sect. \ref{sect:NoiseEstimation}.  Since
the noise depends on the density $D$, it is important to take care of the
density variation throughout the spectrum. We thus scale the simulated spectra
with a scaling factor $c$ such that the desired mean $S/N$ in $B_J$ is
achieved, when the appropriate amount of pixel-wise Gaussian noise is added.
The mean $S/N$ of the scaled spectrum is:
\begin{equation}\label{sn_neu}
  \overline{\left(\frac{S}{N}\right)}_{\mbox{\scriptsize new}}
  =\frac{1}{n}\sum_{i=1}^{n}\frac{c\cdot D_i}{
    a_0+a_1\cdot c\cdot D_i+a_2\cdot c^2\cdot D_i^2},
\end{equation}
We use one set of typical noise coefficients $a_0\dots a_2$ for our
simulations. The appropriate scaling factor $c$ is determined numerically from
Eq. (\ref{sn_neu}) using the Newton-Raphson method. A comparison of a
simulated DA spectrum with a real HES spectrum is shown in Fig.
\ref{slit2objprismdemo}.

\section{Selection of DA white dwarfs}\label{Sect:DAsel}

\begin{figure*}[htbp]
  \begin{center}
    \leavevmode
    \epsfig{file=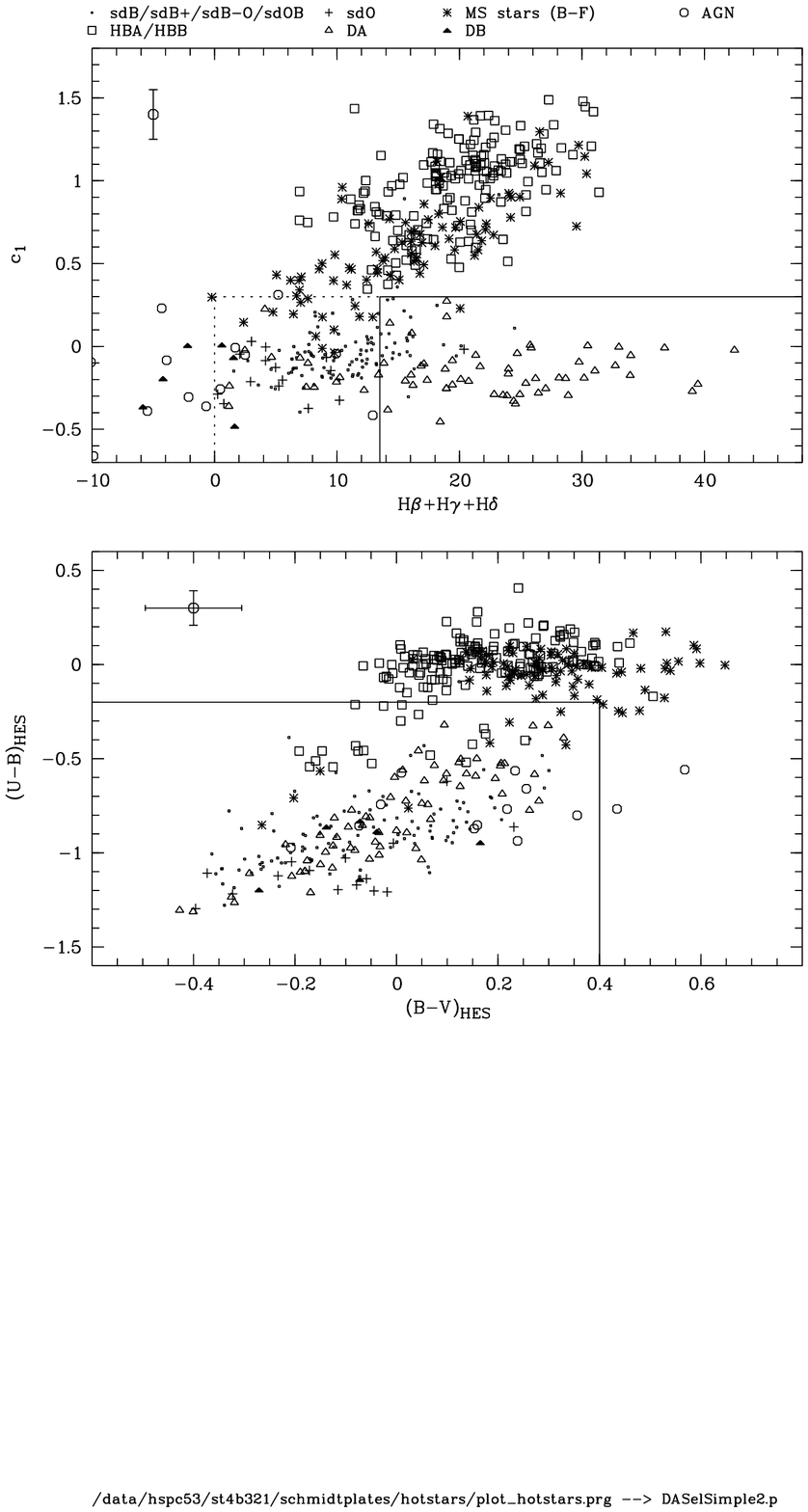, clip=, width=13.5cm,
      bbllx=68, bblly=270, bburx=455, bbury=760}
    \caption{\label{DASelSimple}
      DA selection in colour-colour space and $c_1$ versus Balmer line
      sum feature space. Solid lines: adopted selection box; dashed lines:
      additional parameter area included for the selection of the UVES
      sample. In the upper left corner, error bars for $c_1$, $U-B$ and
      $B-V$ are displayed.}  
  \end{center}
\end{figure*}

Since we are aiming at a very \emph{efficient} selection of white dwarf
candidates, we investigated to what extent different types of hot stars can be
distinguished in the HES in a two-colour diagram ($U-B$ versus $B-V$), and in
the two-dimensional feature space $c_1$ versus {\tt balmsum}. We also explored
several other parameter combinations, including e.g. continuum shape
parameters determined by principal component analysis, but found them to be
less appropriate.  Using various catalogs, we identified 521 hot stars in the
HES. The catalogs are: \citep{Kilkennyetal:1997}, \cite{Wilhelmetal:1999b}, an
updated version of the \cite{Kilkennyetal:1988} subdwarf catalog,
and \cite{McCook/Sion:1999}.  Additionally, 39 HES A-type stars
with known stellar parameters were included.
% Since $U-B$, and especially
% $c_1$ can be easily confused by overlaps, and our colour calibrations are not
% valid for saturated stars, we excluded stars above the saturation threshold,
% and we applied a harder rejection criterion for overlaps (i.e., \emph{no}
% overlapping object detected, instead of allowing for overlapping objects at
% $x>3000$, corresponding to $\lambda<3830$\,{\AA}).

It turned out that high gravity stars (WDs, sdBs, and sdOs) can be
distinguished quite reliably from lower gravity stars (main sequence and
horizontal branch stars). However, the separation between hot subdwarfs and
WDs is much more difficult, since they occupy similar regions in the parameter
space (see Fig. \ref{DASelSimple}). The only way to obtain clean white dwarf
samples is to sacrifice completeness, as we will see below. Note however that
unlike in the EC, it is possible in the HES to compile candidate samples that
have \emph{no} contamination with main sequence or horizontal branch stars.

With the spectral features automatically detected in HES spectra up to now, it
is not yet possible to select DB white dwarfs; in order to do this, Helium
lines would have to be added to the line list. This is currently being done.

As a first step in developing selection criteria for DAs, selection boxes in
the $B-V$ versus $U-B$ and $c_1$ versus Balmer line sum parameter spaces were
chosen such that all DAs were included (see Fig.  \ref{DASelSimple}). A first
set of 47 HES DA candidates (hereafter referred to as the ``UVES sample''; see
Tab. \ref{tab:UVESsample}) selected in this way (and 72 additional WDs)
were observed with VLT-UT2 and UVES between April 4 and June 6,
2000. This set includes seven re-discovered DAs listed in
\cite{McCook/Sion:1999}. Based on pipeline-reduced spectra, 19 of these were
classified as DAs, 26 as hot subdwarfs (e.g., sdB, sdOB, sdO), two as DOs, and
\emph{none} as main sequence or horizontal branch star.

In order to decrease the contamination with hot subdwarfs, we examined if
``harder'' selection criteria could lead to reduced, but acceptable
completeness with respect to DA white dwarfs on the one hand, and a cleaner
candidate sample on the other hand.
\begin{figure}[htbp]
  \begin{center}
    \leavevmode
    \epsfig{file=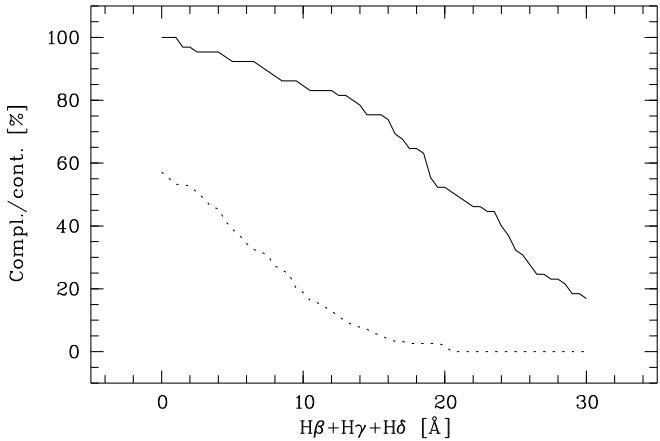, clip=, width=8.8cm,
      bbllx=72, bblly=370, bburx=370, bbury=569}
    \caption{\label{DASelEval} Completeness (solid line) and contamination
    (dotted line) of the DA candidate sample as a function of Balmer line
    equivalent sum.}  
  \end{center}
\end{figure}
Contamination and completeness in dependence of the cutoff in Balmer line
equivalent width sum was evaluated by using the subsample of those 200 stars
from the learning sample of 521 stars mentioned above that were located in the
selection boxes also applied to the UVES sample. The relative numbers of stars
were scaled such that the fraction of hot subdwarfs
and WDs did reflect the fractions in the UVES sample. As can be seen from the
results displayed in Fig. \ref{DASelEval}, it is possible to compile a 80\,\%
complete sample of DA white dwarfs, which is contamined by only 8\,\% hot
subdwarfs, if one confines the sample to objects with $W_\lambda(\mbox{H}\beta
+\mbox{H}\gamma +\mbox{H}\delta) > 13.5$\,{\AA}.

\begin{figure}[htbp]
  \begin{center}
    \leavevmode
    \epsfig{file=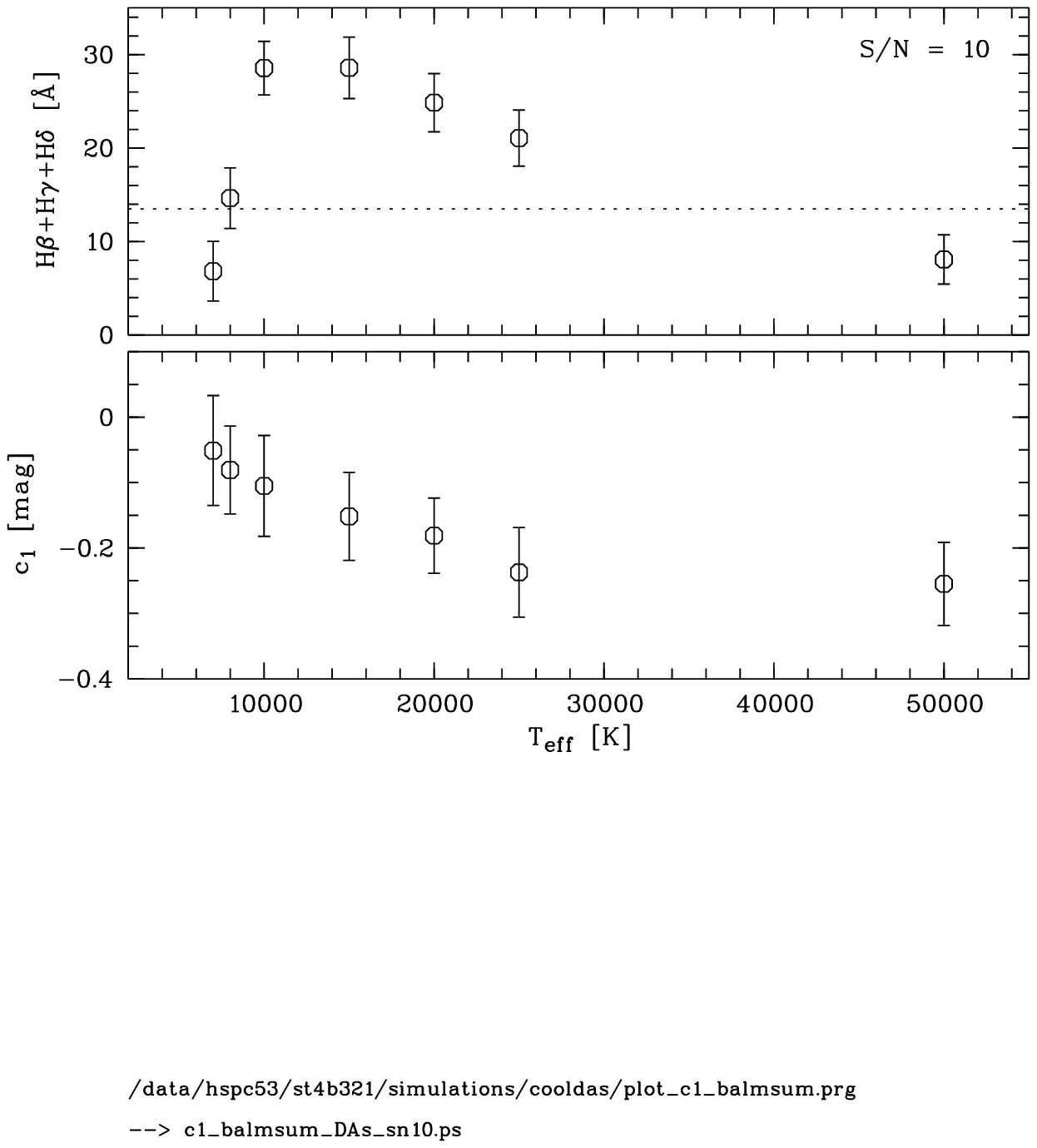, clip=, width=8.8cm,
      bbllx=68, bblly=265, bburx=428, bbury=513}
    \caption{\label{c1_balmsum_DAs_sn10} Balmer line equivalent width sum and
    Stroemgren $c_1$ as a function of effective temperature, determined from
    simulated HES spectra with $S/N=10$ (typically corresponding to
    $B_J=16.4$). In the upper panel the {\tt balmsum}-cutoff adopted in the
    HES is displayed (dotted line); the cutoff in $c_1$ (lower panel) is
    outside of the plotted parameter range. The error bars for $c_1$ refer to
    \emph{internal} errors only.}  
  \end{center}
\end{figure}
Selecting only the objects with very strong Balmer lines leads to
incompleteness at the \emph{hot} end of the DA sample (see Fig.
\ref{c1_balmsum_DAs_sn10}), whereas DAs at
$11\,000\,\mbox{K}>T_{\mbox{\scriptsize eff}}>35\,000\,\mbox{K}$ are very
likely to be selected even at faint magnitudes (that is, at low $S/N$).

The final criteria,
\begin{eqnarray*}
W_\lambda(\mbox{H}\beta+\mbox{H}\gamma +\mbox{H}\delta) &>& 13.5\,\mbox{\AA} \\
c_1 &<& 0.3\\
B-V &<& 0.4\\
U-B &<& -0.2,
\end{eqnarray*}
were \emph{a posteriori} applied to the UVES sample for an independent test.
The resulting sample consists of 15 DAs only; that is, the completeness is
78.9\,\%, which is very close to what was predicted. In a sample of 15
objects, 1.2 subdwarfs are expected if the contamination is 8\,\%, but none
was found, which is consistent with the prediction.

We evaluated the \emph{raw candidate} selection by visually inspecting a
sample of 675 DA candidates at the computer screen. Visual inspection is a
mandatory step in the HES in order to identify plate artifacts, the few
overlaps that are not recognized by the automatic detection algorithm, and to
reject mis-classified and noisy spectra. The 675 candidates were selected from
the 1\,789\,792 overlap-free spectra with $S/N>5$ present on 225 HES plates.
We define the \emph{specifity} of a selection as the fraction of candidate
spectra ($N_{\mbox{\scriptsize can}}$) drawn from the original set of spectra
($N_{\mbox{\scriptsize total}}$); i.e.,
\begin{equation}
 \mbox{specifity}=\frac{N_{\mbox{\scriptsize can}}}{N_{\mbox{\scriptsize total}}}.
\end{equation}
We define the \emph{efficiency} of a selection as the fraction of
\emph{correct} candidates among the automatically selected candidates, as
judged from visual inspection of the objective-prism spectra. The results are
listed in Tab. \ref{tab:sel_eff}.

\begin{table}[htbp]
    \caption{\label{tab:sel_eff}
      Selection specifity and efficiency for DA white dwarfs in the HES,
      determined by visual inspection of candidates from 225 HES fields.
      For the definition of specifity see text.      
      }
  \begin{flushleft}
    \begin{tabular}{lr}\hline\hline
    Specifity  & 1/2652\rule{0.0ex}{2.3ex}\\
    DA candidates                              & 71.6\,\% \\
    Uncertain DA candidates                    &  3.3\,\% \\
    main sequence stars (F or later)           &  7.4\,\% \\
    Subdwarf B stars                           &  5.0\,\% \\
    Horizontal branch A- or B-type stars       &  0.9\,\% \\
    Quasar candidates                          &  0.9\,\% \\
    Noisy spectra (no classification possible) &  4.7\,\% \\
    Spectra disturbed by artifacts             &  4.1\,\% \\
    Overlaps                                   &  2.1\,\% \\\hline\hline
    \end{tabular}    
  \end{flushleft}
\end{table}

\section{Discussion}

The selection method described above allows a very efficient selection of DAs,
with a high completeness ($\sim 80\,\%$). However, this completeness estimate
is based on a sample of objects that are part of the current HES data base of
spectra. It is expected that a significant fraction of WDs are not part of
that data base due to their proper motion (p.m.).  This is because the input
catalog for extraction of objective-prism spectra is generated by using the
DSS~I. Large proper motions and/or large epoch differences between HES and
DSS~I plates (13.5 years on average) currently result in a mis-extraction of
spectra of objects having $\mu_{\alpha}\cdot\Delta\,t_{\mbox{\scriptsize
    HES-DSS~I}}\gtrsim 4''$ (i.e., $\gtrsim3$ pixels), because the spectrum of
the object is looked for at the wrong place. Proper motion in direction of
$\delta$ leads to an offset of the wavelength calibration zero point,
resulting in wrong equivalent width measurements (i.e., the measured widths
are too low).

How large the incompleteness due to the epoch-difference problem is can
roughly be estimated from a cross-identification of the HES with the catalog
of \cite{McCook/Sion:1999}. It lists 2\,187 objects. 1\,633 have an available
$V$ measurement. Of these, 1\,295 (or $79.3$\,\%) lie in the HES magnitude
range ($13\lesssim V\lesssim 17.5$, assuming an average $B-V$ of zero). 606
objects are located in the southern hemisphere ($\delta < \decdegmm{2}{5}$),
and at high galactic latitude ($|b|>30^{\circ}$). Assuming that $79.3$\,\% of
these are in the HES magnitude range, we expect $\sim 480$ WDs to be in the
HES area, and detectable in the HES. Taking into account a loss of $25$\,\%
due to overlapping spectra, we expect 360 known WDs to be found on all 380 HES
plates, and 310 WDs on the 329 plates used so far for the exploitation of the
stellar content. However, in a cross-identification procedure using a
$10''\times 10''$ search box, to compensate for the sometimes very inaccurate
coordinates of \cite{McCook/Sion:1999}, only 151 WDs ($\sim 50$\,\% of the
expected 310) were found.

At the chosen search box width a ``saturation'' of identified objects was
reached; by using a larger box, no further WDs were found.  However, the
experience from an identification of WDs from \cite{McCook/Sion:1999} in the
northern hemisphere sister project of the HES, the Hamburg Quasar Survey
\citep[HQS;][]{Hagenetal:1995}, is that 30--40\,\% of the objects were
identified at positions more than $10''$ away from the nominal position; in
many cases \emph{much} more (up to several arc\emph{minutes}), which explains
why the identification saturation sets in already in for a $10''\times 10''$
search box.

Furthermore, we suspect that two effects may lead to an underestimate of the
completeness of the present DA selection in the HES: First, many WDs
listed in \cite{McCook/Sion:1999} were discovered in the course of proper
motion surveys. Therefore, the catalog is very likely to be biased towards
high p.m. objects, which makes the listed WDs less likely to be detected in
the HES than the real population of WDs. Second, the peak of the magnitude
distribution of the WDs from \cite{McCook/Sion:1999} is more than one
magnitude brighter than that of the HES, so that on average \emph{closer} WDs
(with on average higher p.m.) are listed in \cite{McCook/Sion:1999}.

We conclude that the true completeness of our selection is probably higher
than indicated by the simple cross-comparison.  We conservatively estimate the
proper motion-induced losses of DAs in the HES to be not higher than $\sim
20$\,\%. Special techniques to recover also high p.m. DAs are currently under
development. Such techniques will permit us to construct flux-limited samples
of DA white dwarfs with a very high degree of completeness, useful e.g.\ to
determine luminosity functions.

Another problem of WD selection in the HES might currently result from the
fact that in the HES stellar feature detection algorithm the width of
the absorption lines is held fixed at the measured width of the instrumental
profile, in order to make the algorithm more robust. However, unlike in
``normal'' stars, the absorption lines of WDs are so broad that they are
usually resolved in the HES spectra. Therefore, the equivalent widths of their
absorption lines are expected to be underestimated.  Apart from more
accurately measured equivalent widths, adding the line widths to the
parameters to be fitted would result in the opportunity to employ a further
criterion for the separation between hot subdwarfs and WDs: if the absorption
lines in the spectrum of an object are significantly broader than the
instrumental profile, the object is very likely to be a WD. A modification of
the feature detection algorithm is in progress.

A further improvement of the selection can likely be achieved if instead of
straight lines higher-order selection boundaries are used, as is the case
e.g. in quadratic discriminant analysis. However, defining such selection
boundaries requires \emph{large}, and \emph{well-defined} learning
samples. A lack of model spectra for early-type stars (main sequence as well
as horizontal branch stars) so far prevented us from using a learning sample
of simulated objective-prism spectra for this purpose.

\section{Conclusions}

We described quantitative procedures for the selection of DA white dwarfs in
the digital data base of the HES. Algorithms for the detection of stellar
emission and absorption lines, broad- and intermediate-band colours were
developed. These are not only used for the selection of DAs, but also for a
couple of other interesting stars, e.g. metal-poor stars, and carbon stars.
Simulation techniques allow us to convert model spectra to HES spectra,
which is important for the development of selection criteria, for the
compilation of learning samples used for automatic spectral classification,
and for the determination of selection functions.

DAs can be selected very efficiently in the HES, as required by a currently
ongoing \emph{Large Programme} using VLT-UT2 and UVES, aiming at the detection
of \emph{double degenerates}. A first set of 440 DA candidates not listed in
the catalog of \cite{McCook/Sion:1999} was identified on 225 HES plates (i.e.,
59\,\% of the survey). 

We note that the UVES sample includes a double-lined DA$+$DA binary, and two
further DAs with significant RV variations. The results will be reported
elsewhere. Further investigation of these systems is in progress.

For investigations requiring \emph{complete} samples of DAs (or other objects
having high proper motions), special extraction techniques have to be employed
to take into account the epoch differences between HES plates and the DSS~I,
since the latter is used for object detection in the HES. Such techniques are
currently under development.

\begin{acknowledgements}
  
  N.C. and D.H. acknowledge financial support from Deutsche
  Forschungsgemeinschaft under grants Re~353/40 and Ko~738/10-3, respectively.
  We thank V. Beckmann for observing a part of the DAs that were used to
  determine the HES spectral sensitivity curves.  Precise, photoelectric $UBV$
  and Str\"omgren $uvby$ photometry for HK survey stars was kindly provided by
  T. Beers before publication. We thank D.  O'Donoghue for making EC survey
  photometry in digital form available to us.  G. Pizarro kindly compiled a
  list of HES plate batches from the notes he made at the ESO Schmidt
  telescope. We thank the Paranal staff for carrying out the observations for
  the ESO Large Programme 165.H-0588 in Service Mode at VLT-UT2.

\end{acknowledgements}

\bibliography{classification,datanaly,HES,mphs,ncpublications,ncastro,quasar,statistics,wd}
\bibliographystyle{aabib99}

\begin{appendix}

  \section{The UVES sample}\label{sect:UVESsample}

  In Tab. \ref{tab:UVESsample} we list coordinates, magnitudes, colours,
  Balmer line equivalent widths sums and classifications of the UVES sample of
  47 DA candidates.

\begin{table*}[htbp]
  \caption{\label{tab:UVESsample} The UVES sample, with exception of the DA+DA binary, which
  will be published elswhere. Coordinates are for equinox $2000.0$. In the
  column ``Type'' we list object classifications based on UVES spectra. }
  \begin{center}
    \begin{tabular}{lllccrrrll}\hline\hline
 HE name & R.A. & dec & $B$ & $U-B$ & \multicolumn{1}{c}{$B-V$} & \multicolumn{1}{c}{$c_1$}
 & {\tt balmsum} & Type & McCook \& Sion \rule{0.0ex}{2.3ex}\\\hline
 HE 0952$+$0227 & 09 55 34.6 & $+$02 12 48 & 14.7 & $-1.04$ & $-0.17$ & $-0.10$ &  6.6 & sdO            & \rule{0.0ex}{2.3ex}\\
 HE 0956$+$0201 & 09 58 50.4 & $+$01 47 23 & 15.6 & $-0.64$ & $ 0.06$ & $-0.09$ & 24.7 & DA             & WD 0956$+$020 (DA3)\\
 HE 0958$-$1151 & 10 00 42.6 & $-$12 05 59 & 13.7 & $-1.05$ & $-0.11$ & $-0.15$ &  2.9 & DAB            & \\
 HE 1008$-$1757 & 10 10 33.4 & $-$18 11 48 & 14.9 & $-0.91$ & $-0.15$ & $-0.02$ &  2.5 & DAO (sdO hot)  & WD 1008$-$179 (DA)\\
 HE 1012$-$0049 & 10 15 11.7 & $-$01 04 17 & 15.6 & $-0.86$ & $ 0.06$ & $-0.07$ & 20.5 & DA             & \\
 HE 1026$+$0014 & 10 28 34.8 & $-$00 00 29 & 13.9 & $-0.49$ & $ 0.30$ & $-0.05$ & 23.1 & DA+dM          & WD 1026$+$002 (DA3)\\
 HE 1033$-$2353 & 10 36 07.2 & $-$24 08 34 & 16.0 & $-1.08$ & $-0.34$ & $-0.03$ & 12.4 & sdB (sdOB)     & \\
 HE 1038$-$2326 & 10 40 36.9 & $-$23 42 39 & 15.9 & $-0.86$ & $-0.13$ &  $0.00$ & 10.7 & sdB+cool star? & \\
 HE 1047$-$0436 & 10 50 26.9 & $-$04 52 36 & 14.7 & $-0.96$ & $-0.21$ & $-0.01$ & 10.0 & sdB            & \\
 HE 1047$-$0637 & 10 50 28.7 & $-$06 53 26 & 14.3 & $-1.04$ & $-0.17$ & $-0.11$ &  6.0 & sdO            & \\
 HE 1053$-$0914 & 10 55 45.4 & $-$09 30 58 & 16.4 & $-0.74$ & $ 0.07$ & $-0.12$ & 30.3 & DA             & \\
 HE 1059$-$2735 & 11 01 24.9 & $-$27 51 42 & 15.1 & $-1.23$ & $-0.35$ & $-0.19$ & 10.6 & sdB (sdOB)     & \\
 HE 1117$-$0222 & 11 19 34.7 & $-$02 39 05 & 14.3 & $-0.51$ & $ 0.22$ & $-0.10$ & 24.8 & DA             & \\
 HE 1130$-$0620 & 11 32 41.5 & $-$06 36 53 & 15.8 & $-1.08$ & $-0.09$ & $-0.01$ &  5.3 & sdB (sdOB)     & \\
 HE 1136$-$2504 & 11 39 10.2 & $-$25 20 55 & 13.8 & $-1.09$ & $-0.17$ & $-0.15$ &  9.5 & sdO            & \\
 HE 1142$-$2311 & 11 44 50.2 & $-$23 28 18 & 15.4 & $-1.28$ & $-0.34$ & $-0.23$ &  3.6 & sdO            & \\
 HE 1152$-$1244 & 11 54 34.9 & $-$13 01 17 & 15.8 & $-0.47$ & $ 0.08$ & $-0.13$ & 28.4 & DA             & \\
 HE 1200$-$1924 & 12 02 40.1 & $-$19 41 08 & 14.3 & $-0.95$ & $ 0.21$ & $-0.26$ &  3.0 & sdO            & \\
 HE 1204$-$3217 & 12 06 47.7 & $-$32 34 32 & 15.7 & $-0.84$ & $-0.08$ & $-0.21$ & 27.5 & DA             & WD 1204$-$322 (DA)\\
 HE 1221$-$2618 & 12 24 32.7 & $-$26 35 16 & 14.7 & $-0.79$ & $ 0.07$ & $-0.13$ &  6.7 & sdB+cool star  & \\
 HE 1225$+$0038 & 12 28 07.8 & $+$00 22 17 & 15.2 & $-0.45$ & $ 0.34$ & $-0.10$ & 22.5 & DA             & \\
 HE 1225$-$0758 & 12 27 47.5 & $-$08 14 38 & 14.6 & $-0.85$ & $ 0.06$ & $-0.09$ &  1.8 & DAB, DBA?      & WD 1225$-$079 (DZA)\\
 HE 1229$-$0115 & 12 31 34.8 & $-$01 32 09 & 13.9 & $-0.72$ & $ 0.11$ & $-0.08$ & 24.6 & DA             & WD 1229$-$012 (DA4)\\
 HE 1237$-$1408 & 12 39 56.5 & $-$14 24 48 & 16.1 & $-1.13$ & $-0.07$ & $-0.19$ & 12.0 & sdOB           & \\
 HE 1238$-$1745 & 12 41 01.1 & $-$18 01 58 & 14.3 & $-0.95$ & $ 0.11$ & $-0.16$ &  6.1 & sdO            & \\
 HE 1247$-$1738 & 12 50 22.2 & $-$17 54 47 & 16.2 & $-0.77$ & $ 0.26$ & $-0.35$ & 24.6 & DA+dM          & WD 1247$-$176 (DA)\\
 HE 1256$-$2738 & 12 59 01.4 & $-$27 54 19 & 16.1 & $-1.30$ & $-0.31$ & $-0.36$ &  4.7 & sdO            & \\
 HE 1257$-$2021 & 13 00 27.2 & $-$20 37 27 & 16.5 & $-1.08$ & $ 0.06$ & $-0.12$ &  2.8 & DO/PG1159??    & \\
 HE 1258$+$0113 & 13 00 59.2 & $+$00 57 10 & 16.5 & $-0.95$ & $-0.01$ & $ 0.00$ &  4.2 & sdO            & \\
 HE 1258$+$0123 & 13 01 10.5 & $+$01 07 39 & 16.5 & $-0.24$ & $ 0.24$ & $ 0.26$ & 31.9 & DA             & \\
 HE 1309$-$1102 & 13 12 02.3 & $-$11 18 16 & 16.1 & $-0.80$ & $-0.17$ & $ 0.03$ & 11.4 & sdB            & \\
 HE 1310$-$2733 & 13 12 50.6 & $-$27 49 02 & 14.3 & $-1.23$ & $-0.25$ & $-0.28$ &  8.2 & sdO            & \\
 HE 1314$+$0018 & 13 17 24.7 & $+$00 02 36 & 15.6 & $-1.12$ &  $0.00$ & $-0.23$ & 10.8 & DO             & \\
 HE 1315$-$1105 & 13 17 47.4 & $-$11 21 05 & 15.8 & $-0.42$ & $ 0.23$ & $ 0.05$ & 23.3 & DA             & \\
 HE 1318$-$2111 & 13 21 15.6 & $-$21 27 18 & 14.6 & $-0.99$ & $-0.10$ & $-0.14$ &  4.6 & sdOB           & \\
 HE 1325$-$0854 & 13 28 23.9 & $-$09 09 53 & 15.2 & $-0.64$ & $ 0.05$ & $ 0.04$ & 27.7 & DA             & \\
 HE 1333$-$0622 & 13 36 19.7 & $-$06 37 59 & 16.2 & $-0.62$ & $ 0.15$ & $ 0.11$ & 20.6 & DA+dM          & \\
 HE 1349$-$2320 & 13 52 15.0 & $-$23 34 57 & 15.1 & $-1.20$ & $-0.04$ & $-0.33$ & 10.2 & sdO            & \\
 HE 1352$-$1827 & 13 55 26.6 & $-$18 42 09 & 16.0 & $-0.96$ & $-0.17$ & $-0.06$ & 11.3 & sdB+cool star? & \\
 HE 1355$-$0622 & 13 57 54.3 & $-$06 37 32 & 13.4 & $-1.12$ & $-0.23$ & $-0.05$ &  2.1 & sdO            & \\
 HE 1356$-$1613 & 13 59 12.5 & $-$16 28 01 & 16.1 & $-1.21$ & $-0.35$ & $-0.23$ &  7.9 & sdO            & \\
% HE 1414$-$0848 & 14 16 52.0 & $-$09 02 03 & 16.2 & $-0.52$ & $ 0.09$ & $ 0.14$ & 25.7 & DA+DA          & \\
 HE 1419$-$1205 & 14 22 02.1 & $-$12 19 30 & 16.2 & $-0.98$ & $-0.30$ & $-0.06$ & 11.3 & sdB (sdOB)     & \\
 HE 1502$-$1019 & 15 05 22.7 & $-$10 31 26 & 15.6 & $-1.04$ & $-0.17$ & $-0.25$ &  7.3 & sdOB+cool star & \\
 HE 1511$-$0448 & 15 14 12.9 & $-$04 59 33 & 15.3 & $-1.14$ & $-0.32$ & $-0.05$ &  6.8 & DA             & \\
 HE 1511$-$1103 & 15 14 17.0 & $-$11 14 13 & 14.7 & $-1.12$ & $-0.25$ & $-0.11$ &  1.7 & sdO            & \\
 HE 1512$-$0331 & 15 14 50.1 & $-$03 42 50 & 16.0 & $-1.22$ & $-0.32$ & $-0.35$ &  0.8 & sdO/DAO        & \\\hline\hline
    \end{tabular}
  \end{center}
\end{table*}

  \section{Spectral atlas of DA and DB white dwarfs}\label{Sect:SpectralAtlas}

\begin{figure*}[h]
    \begin{center}
      \leavevmode
      \epsfig{file=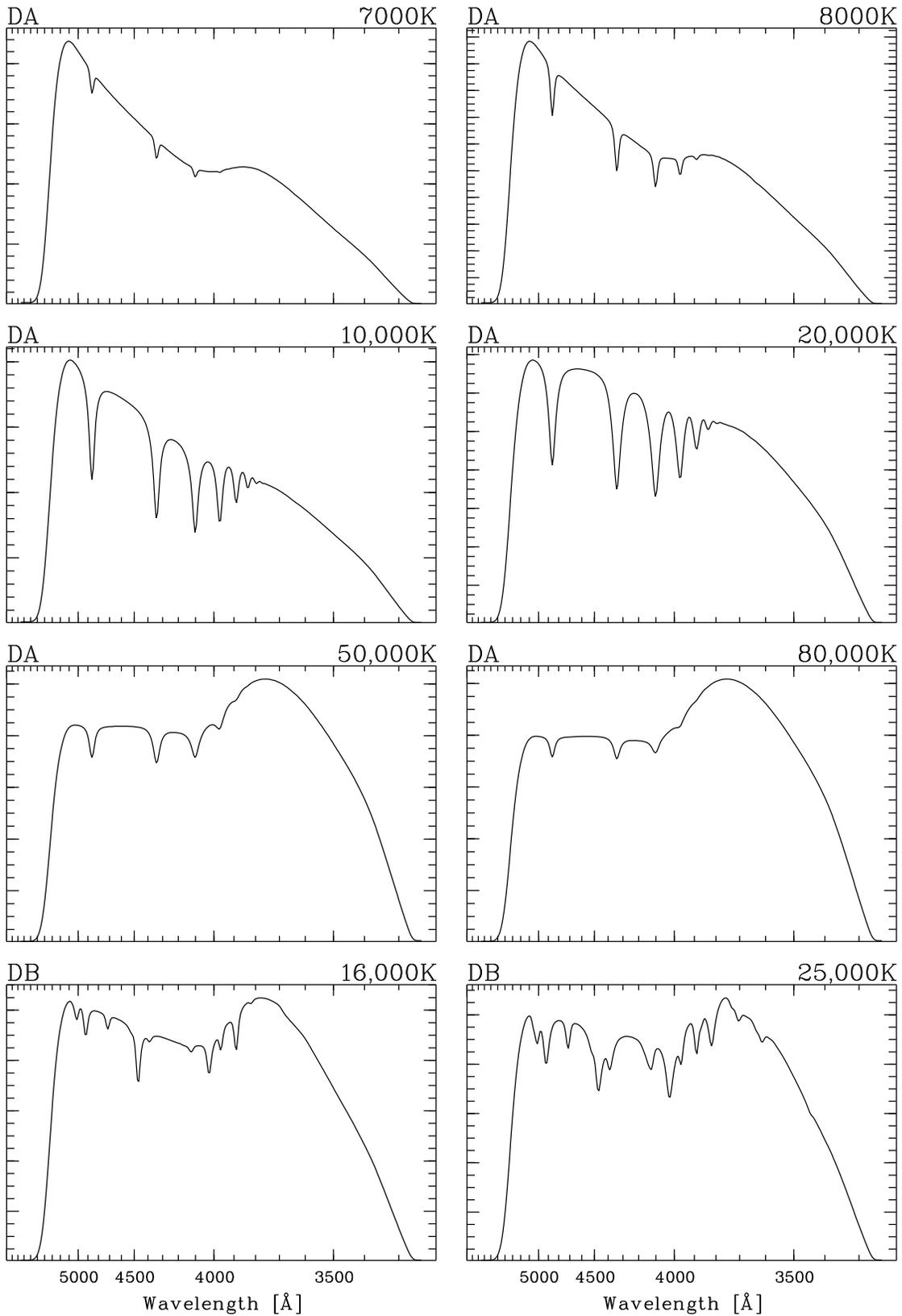, clip=, width=15cm,
        bbllx=80, bblly=125, bburx=500, bbury=735}
      \caption{\label{WDmodels} DA and DB white dwarf model spectra, converted to
        objective-prism spectra. Note that to these spectra no artificial noise
        has been added. }  
    \end{center}
  \end{figure*}

  In Fig. \ref{WDmodels} we display model spectra of DAs and DBs, converted to
  objective-prism spectra with the method described in
  Sect. \ref{slit2objprism}. The model spectra were computed by using
  $\log g=8.0$.  
  
\end{appendix}

\end{document}